\documentclass[useAMS,usenatbib,preprintnumbers,nofootinbib,notitlepage,twocolumn]{revtex4-1}

\usepackage{epsfig}
\input psfig.sty
\usepackage{graphicx}
\usepackage{dcolumn}
\usepackage{bm}
\usepackage{amsmath} 
\usepackage{graphicx}
\usepackage[english]{babel}
\usepackage{mathrsfs}
\usepackage{amsfonts}
\usepackage{amssymb}
\usepackage[latin1]{inputenc}
\usepackage{hyperref}
\usepackage{longtable}
\usepackage{float}
\usepackage{dcolumn}
\usepackage{appendix}
\usepackage{multirow}
\usepackage{color}

\newcommand{\nk}{\textbf{k}}
\newcommand{\nq}{\textbf{q}}
\newcommand{\dphi}{\delta \phi}
\newcommand{\x}{\textbf{x}}

\newcommand{\bra}{\langle}
\newcommand{\ket}{\rangle}
\newcommand{\mH}{\mathcal{H}}
\newcommand{\mP}{\mathcal{P}}
\newcommand{\mR}{\mathcal{R}}

\newcommand{\mO}{\mathcal{O}}
\newcommand{\mF}{\mathcal{F}}
\newcommand{\nn}{\nonumber \\}

\newcommand{\di}{\diamond}  
\newcommand{\eidi}{\epsilon_{1 \diamond}}  
\newcommand{\eiidi}{\epsilon_{2 \diamond}}

\newcommand{\RI}{\text{R,I}}

\begin{document}

\title{Generation of inflationary perturbations in the continuous spontaneous localization model: The second order power spectrum}

\author{Gabriel Le\'{o}n}
\email{gleon@fcaglp.unlp.edu.ar }
\affiliation{Grupo de Astrof\'{\i}sica, Relatividad y Cosmolog\'{\i}a, Facultad
	de Ciencias Astron\'{o}micas y Geof\'{\i}sicas, Universidad Nacional de La
	Plata, Paseo del Bosque S/N 1900 La Plata, Argentina.\\
	CONICET, Godoy Cruz 2290, 1425 Ciudad Aut\'onoma de Buenos Aires, Argentina. } 
	
\author{Mar\'ia P\'ia Piccirilli }
\email{mpp@fcaglp.unlp.edu.ar}
\affiliation{Grupo de Astrof\'{\i}sica, Relatividad y Cosmolog\'{\i}a, Facultad
	de Ciencias Astron\'{o}micas y Geof\'{\i}sicas, Universidad Nacional de La
	Plata, Paseo del Bosque S/N 1900 La Plata, Argentina.\\
	CONICET, Godoy Cruz 2290, 1425 Ciudad Aut\'onoma de Buenos Aires, Argentina. }

\begin{abstract}

Cosmic inflation, which describes an accelerated expansion of the early Universe, yields the most successful predictions regarding  temperature anisotropies in the cosmic microwave background (CMB). Nevertheless, the precise origin of the primordial perturbations  and their quantum-to-classical transition is still an open issue. The continuous spontaneous localization model
(CSL), in the cosmological context, might be used to provide a solution to the mentioned puzzles by considering an objective reduction of the inflaton wave function. In this work, we calculate the primordial power spectrum at the next leading order in the Hubble flow functions that results from applying the CSL model to slow roll inflation within the semiclassical gravity framework.  We employ the method known as uniform approximation along with a second order expansion in the Hubble flow functions.  We analyze some features in the  CMB temperature and primordial power spectra that could help to distinguish between the standard prediction and our approach.
 
\end{abstract}


\maketitle

\section{Introduction}

The most recent observational data obtained from the Cosmological Microwave Background (CMB) are consistent with the hypothesis that the early Universe underwent an accelerated expansion \cite{Planck2018_inflation}. The model to describe that epoch, known as inflation, is now considered as an essential part of the concordance $\Lambda$CDM cosmological model. The success of the inflationary scenario is based on its predictive power to yield the initial conditions for all the observed cosmic structure, which are commonly referred to as  primordial perturbations \cite{mukhanov1992}. 

In the most simple inflationary model, the origin of  primordial perturbations is substantially related to quantum vacuum fluctuations of the scalar field driving the accelerated expansion. Here a subtle question arises: How exactly do these quantum fluctuations become actual (classical) inhomogeneities/anistropies? And in particular, How does the standard inflationary model accounts for the
transition from the initially homogeneous and isotropic quantum state (i.e. the vacuum) into a state lacking such symmetries? It is fair to say that the answer to these questions have not been completely settled, and a large amount of literature has been devoted to this subject \cite{jmartin,jmartinPRL,pedrocsl,Shortcomings,Das2013,kiefer,polarski,pintoneto,valentini,goldstein,ashtekar2020}.

The main reason why this debate continues is because it touches on another controversial issue, i.e. the quantum measurement problem. Specifically, in the standard Copenhagen interpretation of Quantum Mechanics (QM), it is an essential requirement to define (or identify) an observer who performs a measurement with some kind of device. However, in the early Universe there are no such entities, and the measurement problem becomes exacerbated \cite{Bell2,CH1,CH2,Shortcomings,jmartin}.\footnote{Sometimes it is argued that it is us--humans--who are the observers with our own astronomical observations. This argument is rebuked, because if that is the case, then the Universe was homogeneous and isotropic until our astronomers started making observations; however, that is impossible because a Universe that is homogeneous and isotropic contains no astronomers. } One of the first attempts to deal with the aforementioned issue was by invoking the decoherence framework \cite{kiefer,polarski}. Although, decoherence can provide a partial understanding of the issue, it does not fully addresses the problem mainly because decoherence does not solve the quantum measurement problem \cite{adler}. We will not dwell in all the conceptual aspects regarding the appeal of decoherence during inflation; instead, we refer the interested reader to Refs.  \cite{Shortcomings,elias2015} for a more in depth analysis.

There are many approaches to the subject of Foundations of Quantum Theory and in particular to the quantum measurement problem, but a good method to classify them is provided by the result of \cite{Maudlin}. There, one can find a particularly useful way to state \textit{the measurement problem}, which consists in a list of three statements that cannot be all true at the same time:

\begin{itemize}
	\item[A.] The physical description given by the quantum state is complete.
	\item[B.] Quantum evolution is always unitary.
	\item[C.] Measurements always yield definite results.
\end{itemize}

The need to forsake (at least) one of the above forces one toward a specific conceptual path depending on the choice one makes. Concretely speaking, forsaking (A)  leads naturally to hidden variable theories, such as de Broglie-Bohm or ``pilot wave'' theory \cite{bohmbook,durrbook}. Forsaking (B), one is naturally led to collapse theories and which for the cosmological case seem to leave no option but those of the spontaneous kind, such as the Ghirardi-Rimini-Weber or Continuous Spontaneous Localization models \cite{grw,per,bassi}. The reason is that there is clearly no role for conscious observers or measuring devices that might be meaningfully brought to bear to the situation at hand. Finally, forsaking (C) seems to be the starting point of approaches such as the Everettian type of interpretations \cite{manyworlds}.  The latter, again, seem quite difficult to be suitably implemented in the context at hand, simply because ``observers'', ``minds'', and such notions, that play an important role in most attempts to characterize the world branching structure in those approaches,  can only be accounted for within a Universe in which structure has already developed, well before the emergence of the said entities.

All of those approaches have been followed to investigate the
generation of primordial perturbations during inflation \cite{pia_tesis,pedrocsl,jmartin,jmartinPRL,Das2013,LB15,pintoneto,goldstein,nomura2011}. In the present work,
we will focus on the Continuous Spontaneous Localization (CSL) model
applied to the standard slow roll inflationary scenario, and just for notation comfort, from now on we will refer to this idea as the CSL inflationary model
(CSLIM). Other applications of the CSL model to cosmology have been analyzed recently, for instance to account for the late-time accelerated expansion of the Universe \cite{sudarskyPRL,cristobal}.

 Several aspects of the CSLIM have been studied before. The
first implementation, based on the semiclassical gravity framework\footnote{It is worthwhile to mention that the CSL model has
also been applied to inflation using the Mukhanov-Sasaki variable, which quantizes both the metric and inflaton perturbations \cite{jmartin,jmartinPRL,Das2013,hinduesT}.  },
was done in \cite{pedrocsl}. Afterwards,  using observational data it was possible to statistically constrain the cosmological parameters of the model; also a Bayesian analysis was performed in order to compare the model performance within the standard cosmological model  \cite{pia_tesis}.

 Moreover, in \cite{lucila,nobmodesshort,nobmodesbig}, and working in the context of
semiclassical gravity, it was shown that the CSLIM predicts a strong suppression of
primordial B-modes, i.e. the predicted amplitude of the tensor power
spectrum is very small generically (undetectable by current
experiments). Also in \cite{eternalcsl} it was found that, when enforcing
the CSLIM, the condition for eternal inflation can be
bypassed.

One of the main features of the CSLIM is that it modifies the standard primordial power spectrum through a characteristic $k$ dependence \cite{pia_tesis}; specifically,  the spectrum is of the form $P(k) \propto (k/k_\diamond)^{n_s-1} C(k)$, where $C(k)$ is a new function of the model's parameters (and $k_\diamond$ is the pivot scale). The predicted spectral index $n_s$ is given in terms (as in the traditional approach) of the slow roll parameters or equivalently in terms of the Hubble flow functions (HFF). At this point, we introduce the main motivation for the present work; our purpose is to answer the question: How can one distinguish the $k$ dependence introduced by the CSLIM from a ``simple'' running of the spectral index? and Is it possible to use observational data (recent or future) to answer that question? Here we remind the reader that the running of the spectral index is traditionally interpreted as an extra $k$ dependence induced, in the power spectrum, by the spectral index  $n_s(k)$. In single field slow roll inflation, one immediately realizes that an attempt to answer those questions requires first a calculation of the power spectrum at second order in the HFF within the CSLIM. In the present paper, we present the result and computational details for such calculation. Furthermore, we perform a comparison between our prediction and  the second order power spectrum given in the traditional approach \cite{venninSR,terreroSR,ringevalSR,liddleSR,lythSR,jmartinSR,jmartinSR2}. Also we perform a preliminary analysis of the observational consequences for each model. Our calculations made use of the uniform approximation method \cite{habib2004,venninSR}; these are supplemented in two Appendices, where one can also find  our prediction for $n_s$ and $\alpha_s$ at higher order in the HFF.

We can further motivate the significance of the sought result in this paper. Recent data from \emph{Planck} collaboration seem to indicate that a scale dependence of the scalar spectral index is still allowed by observations \cite{Planck2018_inflation}. As we have mentioned, this scale dependence of $n_s$ is known as the running of the spectral index $\alpha_s$. The current data from \emph{Planck} indicates that $\alpha_s = -0.0045 \pm  0.0067$ at 68\% CL and $\alpha_s = -0.005 \pm 0.013$ at 95\% CL (when the running of the running of the spectral index is set to zero). Although these values are consistent with a zero running, future experiments may detect a non-zero value of $\alpha_s$. The relevant issue here would be the order of magnitude of $\alpha_s$. 

Let us recall that at the lowest order in the HFF, the standard prediction from slow roll inflation yields: $n_s -1 = -2 \eidi - \eiidi$, $r= 16 \eidi $ (known as the tensor-to-scalar ratio) and $\alpha_s = -2 \eidi \eiidi - \eiidi \epsilon_{3 \di}$, where $\epsilon_{j \di}$ denotes the HFF  evaluated at the pivot scale; consequently, $\alpha_s = (n_s-1 + r/8)(r/8 + \epsilon_{3 \di})$. Furthermore, as more tight constraints on $r$ are obtained by future collaborations [see e.g. \cite{QUBIC}], a plausible scenario could ensue: It may be the case that $r$ would remain undetected, decreasing the order of magnitude of $\epsilon_{1 \diamond}$ allowed by the data. In that case, a conservative estimate for the magnitude of the running would be $|\alpha_s| \simeq |n_s-1| |r/8 + \epsilon_{3 \diamond}| $. However, assuming also a detection of the running of order $|\alpha_s| \simeq 10^{-3}$, and taking into account that current data indicate $|n_s - 1| \simeq 10^{-2}$, then we would have the estimate $|\epsilon_{3 \diamond}| \simeq 10^{-1}$. That result can be puzzling for the traditional slow roll inflationary paradigm, because one would have $ |\epsilon_{3 \diamond}| > |\epsilon_{2 \diamond}| > |\epsilon_{1 \diamond}|$. In other words, the so called hierarchy of the HFF \cite{liddleSR} would be lost, suggesting a possible inconsistency with the single field slow roll inflationary model \cite{runninglewis}. Note that $|\alpha_s| \simeq 10^{-3}$ is not an unrealistic estimate based on the current $1 \sigma$,$2 \sigma$ CL reported by \emph{Planck} \cite{Planck2018_inflation} and by future observations \cite{japoneses21cm}. 

Moreover, a recent theoretical motivated proposal, known as the Trans-Planckian Censorship Conjecture (TCC) \cite{vafa1}, leads to the prediction of a negligible amplitude of primordial gravitational waves, that is $ |\epsilon_{1 \diamond}| < 10^{-31}$ \cite{vafa2}. The TCC simply put states that in an expanding Universe
sub-Planckian quantum fluctuations should remain quantum and can never become larger than the Hubble horizon and classically freeze.\footnote{The TCC serves to address the trans-Planckian problem for cosmological fluctuations \cite{transplanck1,transplanck2,brahma}. In particular, it is conjectured that the trans-Planckian problem can never arise in a consistent theory of quantum gravity and that all models which would lead to such issues are inconsistent and belong to the Swampland.} Furthermore, it has been found \cite{TCCbrandenberger} that a large value of the second slow-roll parameter and a small $\epsilon_{1 \diamond}$ is essentially preferred not only by the TCC, but also by the so called ``swampland conjecture,'' which is more general.  While, we will left for future work how exactly the TCC could be implemented in the CSLIM, the implications of the TCC do serve to highlight that it is not quite improbable that predictions and observations in standard slow roll inflation might face some issues in the future.

The CSLIM also predicts a strong suppression of primordial gravity waves, but in this case the tensor modes are generated by second order scalar perturbations \cite{nobmodesbig,nobmodesshort}; in fact, an estimate for the tensor-to-scalar ratio has been obtained in Ref. \cite{nobmodesshort}: $r= 10^{-7} \epsilon_{1 \diamond}^2$. This result means that in the CSLIM, $r$ is no longer related at the leading order with $n_s-1$ and $\alpha_s$, which contrasts with the standard prediction. Moreover, since in the CSLIM the predicted spectrum has an extra $k$ dependence through the function $C(k)$, then, in principle, it is possible that $C(k)$ acts as a ``running effect'' which does not depend entirely on $\alpha_s$. As a consequence, the supposed scenario above in the traditional approach, and which would lead to inconsistencies in the slow roll inflationary model, might be resolved within the CSLIM. In particular, a non-detection of $r$ (with tightest constraints) and a sufficiently high detection of a running of the spectral index could be consistent within our proposed framework, but the hierarchy of the HFF would not be violated (as would be the case in the standard approach). These plausible sequence of events, would also serve to show that the CSLIM is not ``just a philosophically'' motivated model (as sometimes is often dismissed) but that it can have important observational consequences.

Thus, in the present work, we will make a first step in that direction, obtaining a prediction for the primordial spectrum at second order in the HFF. This will allow us to analyze clearly the dependence on $k$ of the primordial spectrum, i.e. to single out the contribution given by $\alpha_s$ and $C(k)$ in the predicted form of the power spectrum. Hopefully, future observations could be used to perform a full data analysis using the result obtained here. 

The paper is organized as follows: In Sec. \ref{Sec2}, we present the technical setting that, based on the semiclassical gravity framework,  represents an adequate application of the collapse hypothesis to standard slow roll inflation; this is done at second order in the HFF, also we show how we can obtain a formula for the primordial power spectrum with the previous considerations.  In Sec. \ref{Sec3} the quantum treatment of inflaton is shown by taking into account the CSL model, the novel feature here, with respect to previous works, is the second order equations in the HFF. These equations enable us to obtain the primordial spectrum at the next leading order. In Sec. \ref{secPSanalysis}, 
we compare the primordial power spectrum obtained in the previous section with the phenomenological expression from standard inflationary models. Specifically, we plot the primordial
power spectrum at second order for some particular parameterizations of the collapse parameter and compare it with the primordial spectrum preferred by the data, which corresponds to the standard prediction in slow roll inflation. Moreover, we present our prediction for the CMB temperature fluctuation spectrum and show that possible differences exist with respect to the best fit model obtained in traditional slow roll inflation. The analysis presented in this section takes into account the inflation parameters $A_s$, $n_s$ and $\alpha_s$.  Finally, in Sec. \ref{Sec5}, we summarize the main results of the paper and present our conclusions.

We have included two appendixes with the aim to provide supplementary material for the reader interested in all the computational details. Appendix \ref{csl_eq_appendix} 
contains the technical steps required to solve of the CSL equations at second order in the HFF, these are based on the uniform approximation method. Employing those results, in Appendix \ref{app:calculationPs} we provide the calculations used to obtain the primordial power spectrum at second order, and we also include the prediction for the spectral and running spectral indexes at third and fourth order respectively.


Regarding notation and conventions, we will work with signature $(-, +, +, +)$ for the metric, and we will use units where $c = \hbar = 1$ but keep the gravitational constant $G$.

\section{The collapse proposal and the primordial power spectrum}
\label{Sec2}

Before addressing in full detail the main equations of our model, we present the framework that underlies our description of the space-time metric and that of the inflaton \cite{PSS,Shortcomings,alberto,erandy,benito,benito2019}. The proposed model is based on the semiclassical gravity framework, in which gravity is treated classically and the matter fields are treated quantum mechanically.  This approach accepts that gravity is quantum mechanical at the fundamental level, but considers that the characterization of gravity in terms of the metric is only meaningful when the space-time can be considered classical. Therefore, semiclassical gravity can be treated as an effective description of quantum matter fields living on a classical space-time. Clearly this approach is very different from the standard inflationary theory, in which the perturbations of both the metric and the matter fields are treated in quantum mechanical terms. The framework employed is thus based on  semiclassical Einstein's equations (EE),
\begin{equation}\label{scEE}
G_{ab}  = 8 \pi G \bra \hat T_{ab} \ket.
\end{equation}

In our approach the initial state of the quantum field is taken to be the same as the standard one, i.e. the Bunch-Davies (BD) vacuum.  Nonetheless, the self-induced collapse will spontaneously change this initial state into a final one that does not need to share the symmetries of the BD vacuum. These symmetries are homogeneity and isotropy. Consequently, after the collapse, the expectation value $\bra \hat T_{ab} \ket$ will not have the symmetries of the BD vacuum, and this will led, through semiclassical EE, to a geometry that is no longer homogeneous and isotropic generically. The interested reader can consult Refs. \cite{alberto,erandy,benito,benito2019}; in those works the formalism of the collapse proposal within the semiclassical gravity framework has been developed. In the present paper, we will only make use of the most relevant equations.

\subsection{Classical description of the perturbations}
\label{classical}

As in standard slow roll inflationary models, we consider the action
of a single scalar field, minimally coupled to gravity, with an
appropriate potential:
\begin{equation}\label{action0}
S[\phi,g_{ab}] = \int d^4x \sqrt{-g} \bigg[ \frac{1}{16 \pi G} R[g] 
- \frac{1}{2}\nabla_a \phi \nabla_b \phi g^{ab} - V[\phi] \bigg]. 
\end{equation}
The background metric is described by a flat FRW spacetime, with $a(t)$ the scale factor. Meanwhile the matter sector can be modeled by a scalar field which can be decomposed into a homogeneous part plus ``small'' perturbations $\phi(\x,t) = \phi_0(t) + \dphi(\x,t)$. 


In order to describe slow roll (SR) inflation, it is convenient to introduce the Hubble flow functions $\epsilon_i$ (HFF) \cite{jmartinSR}, these are defined as
\begin{equation}\label{defepsilonn}
\epsilon_{n+1} \equiv \frac{d \ln \epsilon_n}{d N}, \qquad \epsilon_0 \equiv \frac{H_{\text{ini}}}{H},
\end{equation}
where $N \equiv \ln (a/a_{\text{ini}}) $ is the number of e-folds from the beginning of inflation; $H \equiv  \dot a/a$ the Hubble parameter and the dot denotes derivative respect to cosmic time $t$. Inflation occurs if $\epsilon_1 < 1$ and the slow roll approximation assumes that all these parameters are small during inflation $|\epsilon_n| \ll 1$. Additionally, since $d N = H dt $, it is straightforward to obtain another useful expression for the HFF, i.e.
\begin{equation}\label{dotepsilon}
\dot \epsilon_n = H \epsilon_n \epsilon_{n+1}.
\end{equation}

In terms of the first two HFF, the dynamical equations for the homogeneous part of the model can be expressed as
\begin{equation}\label{friedamnnSR}
H^2 = \frac{V}{ M_P^2 (3-\epsilon_1)},
\end{equation}
\begin{equation}\label{KGSR}
3H \dot \phi \left(1-\frac{\epsilon_1}{3} + \frac{\epsilon_2}{6} \right) = - \partial_\phi V,
\end{equation}
where $M_P^2 \equiv 1/(8 \pi G)$ is the reduced Planck's mass. The previous equations are exact.

%

%

Let us now focus on the perturbations part of the theory. We start by switching to conformal coordinates; thus, the components of the background metric are $ g_{\mu \nu}^{(0)} = a(\eta) \eta_{\mu \nu}$, with $ \eta $ the conformal cosmological time; $\eta_{\mu \nu}$ the components of the Minkowskian metric. 



We choose to work in the longitudinal gauge; in such a gauge, and focusing on the scalar perturbations at first order,  the line element associated to the metric is:
\begin{equation}
ds^2 = a^2(\eta) \left[ - (1+2\Phi) d\eta^2 + (1-2 \Psi) \delta_{ij} dx^i dx^j \right],
\end{equation}
where $\Phi$ and $\Psi$ are scalar fields, and $i,j = 1,2,3$.  Einstein's equations (EE) at first order in the perturbations,  $\delta G_0^0 = 8  \pi G \delta T_0^0$, $\delta G_i^0 = 8 \pi G \delta T_i^0$ and $\delta G^i_j =  8 \pi G \delta T^i_j$, are given respectively by
\begin{equation}\label{00inf1}
\nabla^2 \Psi -3\mH(\mH\Phi + \Psi') = 4 \pi G [-\phi_0'^2 \Phi + \phi_0' 
\dphi' 
+ \partial_\phi V a^2 \dphi],
\end{equation}
\begin{equation}\label{0iinf1}
\partial_i (\mH \Phi + \Psi') = 4 \pi G \partial_i ( \phi_0' \dphi),
\end{equation}
\begin{eqnarray}\label{ijinf1}
& & [\Psi'' + \mH(2\Psi+\Phi)' + (2\mH' + \mH^2)\Phi + \frac{1}{2} 
\nabla^2 (\Phi - \Psi)] \delta^i_j \nonumber \\
&-&\frac{1}{2} \partial^i \partial_j (\Phi - \Psi) =   4 \pi G 
[\phi_0' \dphi' -\phi_0'^2 \Phi  - \partial_\phi V a^2 
\dphi]\delta^i_j. \nonumber \\
\end{eqnarray}
Equation \eqref{ijinf1} with components $i \neq j$ lead to $\Psi = \Phi$, from now on we will use this result and refer to $\Psi$ as the Newtonian potential. Furthermore, in the longitudinal gauge $\Psi$ represents the curvature perturbation (i.e. the intrinsic spatial curvature on hypersurfaces on constant conformal time for a flat  Universe). Subtracting Eq. \eqref{00inf1} from \eqref{ijinf1}, together with \eqref{0iinf1} and the motion equation for the homogeneous part of the scalar field $a^2 \partial_\phi V = -\phi_0'' - 2 \mH \phi_0'$, one obtains
\begin{equation}\label{psievo}
\Psi'' - \nabla^2 \Psi + 2 \left(\mH - \frac{\phi_0''}{\phi_0'} \right) \Psi' + 2 \left( \mH' - \frac{\mH \phi_0''}{\phi_0'}        \right) \Psi = 0.
\end{equation}
Regarding notation, primes denote derivative with respect to conformal time $\eta$, and $\mH \equiv a'/a$.

Switching to Fourier's space\footnote{We define the Fourier transform of a function $f(\x,\eta)$ as 
	$$f(\x,\eta) = \frac{1}{(2 \pi)^{3/2}} \int_{ \mathbb{R}^3} d^3k \: e^{i \nk \cdot \x} f_\nk (\eta) $$ }, in the super-Hubble limit $k \eta \to 0$, the solution to the above differential equation is
\begin{equation}\label{psisol2}
\Psi_k (\eta) = C_G (k)  [ \epsilon_1 + (\epsilon_1^2 + \epsilon_1 \epsilon_2 ) ] + \mO(\epsilon^3),
\end{equation}
where $C_G(k)$ is a constant fixed by the initial conditions. Also note that solution \eqref{psisol2} is approximately constant.   From \eqref{psisol2} and \eqref{dotepsilon}, it follows that  $\Psi_k'$ is order 2 at the lowest order in the HFF. In particular, we have that $ \Psi_k' = C_G (k) \epsilon_1 \epsilon_2 \mH  + \mO(\epsilon^3) $; hence we approximate 
\begin{equation}\label{psiprima}
\Psi_k' \mH^{-1} \simeq \epsilon_2 \Psi_k .
\end{equation}
This will be a useful result in the following, however note that the approximation breakdowns at order 3 or higher in $\epsilon_n$.

The collapse of the inflaton's wave function, which is governed by the CSL mechanism, is the process that generates the curvature perturbations. We will be more specific in the next section, but for now let assume that the CSL process simply changes randomly the initial state of the field to a different one. This mechanism can be implemented in the early Universe using the semiclassical gravity framework. The semiclassical EE at linear order in the perturbations read $\delta G_{ab} = 8 \pi G \bra \delta \hat T_{ab} \ket$.

Therefore, the semiclassical version of Eq. \eqref{0iinf1} in Fourier's space,  $k_i (  \Psi_\nk + \mH^{-1} \Psi_\nk') = 4 \pi G k_i \mH^{-1}  \phi_0' \bra \hat \dphi_\nk \ket$ together with \eqref{psiprima}, yields
\begin{equation}\label{masterpsi0}
\Psi_\nk + \epsilon_2 \Psi_\nk \simeq 4 \pi G  \mH^{-1}  \phi_0' \bra \hat \dphi_\nk \ket,
\end{equation}
note that \eqref{psiprima} comes from solving \eqref{psievo}, that is equations  $\delta G_0^0 = 8  \pi G \delta T_0^0$, $\delta G_i^0 = 8 \pi G \delta T_i^0$ and $\delta G^i_j =  8 \pi G \delta T^i_j$ have been combined to solve for $\Psi_\nk$. 

We can rewrite Eq. \eqref{masterpsi0} in terms of the HFF only. Taking the derivative of Eq. \eqref{friedamnnSR}  with respect to $t$ and combining it with: Eq. \eqref{KGSR}, the defintion $\epsilon_{1} = - \dot H/H^2$ and $\dot \epsilon_1 = H \epsilon_1 \epsilon_2$; we can find that $\dot \phi_0^2 = M_P^2 H^2 2 \epsilon_1$ or equivalently in conformal coordinates 
\begin{equation}\label{phi0pepsilon}
\phi_0'^2 = M_P^2 \mH^2 2 \epsilon_1,
\end{equation}
that relation is exact. Finally, substituting Eq. \eqref{phi0pepsilon} into \eqref{masterpsi0} leads to the following main equation for the metric perturbation:
\begin{equation}\label{masterpsi}
\Psi_{\nk} \simeq \frac{  1  }{ M_P }  \sqrt{ \frac{\epsilon_1}{2 } } \frac{\bra \hat \dphi_{\nk} \ket }{ (1+\epsilon_2) },
\end{equation}
the approximation is valid up to order 2 in $\epsilon_n$.

This is the main result of the present subsection. Equation \eqref{masterpsi} indicates that when the state is the vacuum, one has $\bra 0 | \hat \dphi_\nk |0  \ket = 0$, i.e. there are no perturbations at any scale $k$; thus $\Psi_\nk =0$. It is only after the collapse has taken place  $| 0 \ket \to | \Xi \ket$, that the expectation value satisfies $\bra \hat \dphi_\nk \ket \neq 0$, and thus giving birth to the primordial perturbations. 

\subsection{The scalar power spectrum}

In this subsection, we want to find an expression for the scalar power spectrum in terms of the metric perturbation equation \eqref{masterpsi}.  We begin by recalling a well-known quantity defined as
\begin{equation}\label{relacionRyPsiexacta}
\mR \equiv \Psi  + \left(  \frac{2 \rho}{3}   \right) \left( \frac{\mH^{-1}  \Psi' + \Psi}{\rho + P}\right),
\end{equation}
where $\rho$ and $P$ are the energy and pressure densities associated to the type of matter driving the expansion of the Universe. The importance of the quantity $\mR$ is that, for adiabatic perturbations, it is conserved for super-Hubble scales, irrespective of the cosmological epoch one is considering. The type of cosmological epoch is characterized by the equation of state $P = \omega \rho$. For a matter dominated epoch $\omega \simeq 0$, and for a radiation dominated epoch $\omega \simeq 1/3$. The Newtonian potential $\Psi$, is also a conserved quantity for super-Hubble scales, but its amplitude changes between epoch transitions; on the contrary, the amplitude of $\mR$ does not change during the transitions. The amplitude variation of $\Psi$ during the transition from  radiation to  matter dominated epoch is not very significant, $|\Psi^{\text{matt.}}| \simeq (9/10) |\Psi^{\text{rad.}}|$. Nevertheless, the amplitude variation between inflation and radiation era does changes significantly; let us see this explicitly. 

During inflation  $\rho + P = \phi_0'^2/a^2 = M_P^2 \mH^2 2 \epsilon_1 /a^2$, and because of Friedmann's equation $\mH^2 = a^2 \rho/ 3 M_P^2$, we have
\begin{equation}
\mR = \Psi + \frac{1}{\epsilon_1}\left( \mH^{-1}  \Psi' + \Psi \right).
\end{equation}
The above equation is exact. However, using approximation \eqref{psiprima} for the Fourier components results in
\begin{equation}\label{relacionRyPsi}
\mR_k \simeq \frac{\Psi_k}{\epsilon_1} (1+ \epsilon_1 + \epsilon_2).
\end{equation}

On the other hand, during the radiation dominated epoch $\mR_k = (3/2) \Psi_k^{\text{rad.}}$. Since $\mR$ is a conserved quantity, hence, we can obtain the change in the amplitude of the Newtonian potential from the inflationary epoch to the radiation dominated epoch,
\begin{equation}\label{Psir}
|\Psi_k^{\text{rad.}}| = \frac{2 (1+ \epsilon_1 + \epsilon_2) }{3 \epsilon_1}  |\Psi_k|.
\end{equation} 
Thus, in the radiation epoch, the amplitude of the Newtonian potential during inflation is amplified by a factor of  $1/\epsilon_1$.

Another important aspect of the quantity $\mR$ is that in the comoving gauge, it represents the curvature perturbation. In fact,  the primordial power spectrum usually shown in the literature is associated to $\mR$. The scalar power spectrum (associated to the curvature perturbation in the comoving gauge) in Fourier space is defined as
\begin{equation}\label{PSdef}
\overline{\mR_{\nk}\mR^*_{\nq}} \equiv \frac{2 \pi^2}{k^3} \mP_{s} (k) \delta(\nk-\nq),
\end{equation} 
where $\mP_{s} (k)$ is the dimensionless power spectrum. The bar appearing in \eqref{PSdef} denotes an ensemble average over possible realizations of the stochastic field $\mR_{\nk}$. In the CSLIM each realization will be associated to a particular realization of the stochastic process characterizing the collapse process.

On the other hand, our main equation from the last subsection \eqref{masterpsi}, was obtained in the longitudinal gauge. Fortunately, Eq. \eqref{relacionRyPsiexacta} relates $\Psi$ and $\mR$  exactly; in other words, we can compute the curvature perturbation in the longitudinal gauge using the CSLIM, and then switch to the comoving gauge in order to compare the primordial spectrum obtained in our model with the standard one. Furthermore, during inflation, we can use approximation \eqref{relacionRyPsi} to compute the scalar power spectrum, associated to $\mR_{\nk}$, that results from our main equation \eqref{masterpsi}. This is, 
\begin{equation}
\overline{\mR_\nk \mR_\nq^*} =  \frac{1}{2 M_P^2 \epsilon_1 }  \frac{(1+ \epsilon_1 + \epsilon_2)^2}{(1+\epsilon_2)^2}  \overline{\bra \hat \dphi_\nk \ket \bra \hat \dphi_\nq \ket^*}.
\end{equation}

Therefore, we can identify the scalar power spectrum as 
\begin{equation}\label{masterPS}
\mP_{s} (k) \delta(\nk-\nq) = \frac{k^3}{4 \pi^2 M_P^2 \epsilon_1 }  \frac{(1+ \epsilon_1 + \epsilon_2)^2}{(1+\epsilon_2)^2}  \overline{\bra \hat \dphi_\nk \ket \bra \hat \dphi_\nq \ket^*}.
\end{equation}

The quantity $\overline{\bra \hat \dphi_\nk \ket \bra \hat \dphi_\nq \ket^*}$, must be evaluated in the super-Hubble regime $k \eta \to 0$. In the next section, we will focus on that quantity.

\section{Quantum treatment of the perturbations: The CSL approach}
\label{Sec3}

We now proceed to describe the quantum theory of the perturbations. Our  treatment is based on the QFT of  $\dphi (\x,\eta)$ in a curved background described by a  quasi--de Sitter spacetime. Expanding the action \eqref{action0} up to second order in the perturbations, one can find the action associated to the matter perturbations. Given that we are working within the semiclassical gravity framework, we are only interested in quantize the matter degrees of freedom. Introducing the rescaled field variable $y=a\dphi$, the second order action is $\delta^{(2)} S =  \int d^4x \delta^{(2)} \mathcal{L}$, where
\begin{eqnarray}\label{action2}
  \delta^{(2)} \mathcal{L} &=& \frac{1}{2} \bigg[  y'^2   - (\nabla y)^2  - y^2 a^2  V_{, \phi \phi} +  \frac{a''}{a} y^2\bigg] \nn
  &+& a [4\phi_0' \Psi' y - 2 a^2 V_{, \phi}  \Psi y  ]
  \end{eqnarray}
and $V_{, \phi}$ indicates partial derivative with respect of $\phi$.
Note that in $\delta^{(2)} \mathcal{L}$ there are terms containing metric perturbations. In the vacuum state, according to our approach, $\Psi = \Psi' =0$. However, since the CSL mechanism is a continuous collapse process, the quantum state characterizing the system will change from $| 0 \ket$ to a new final state $| \Xi \ket$. As a consequence, the metric perturbations (which are always classical) will be changing from zero to a non-vanishing value in a continuous manner. Thus, including the terms containing $\Psi$ and $\Psi'$ in the action can be considered as a backreaction effect of the CSL model, and as we will see this effect is of second order in the HFF.

We next switch to Fourier space. This is justified by the fact that we work with a linear theory and, hence, all the modes evolve independently. We define the field's modes as
\begin{equation}
y(\x,\eta) = \frac{1}{(2 \pi)^{3/2}} \int_{\mathbb{R}^3}  d^3 k \: y_{\nk} (\eta) e^{i \nk \cdot \x},
\end{equation}
\begin{equation}
\Psi(\x,\eta) = \frac{1}{(2 \pi)^{3/2}} \int_{\mathbb{R}^3}  d^3 k \: \Psi_{\nk} (\eta) e^{i \nk \cdot \x},
\end{equation}
with $y_{-\nk} = y_\nk^*$ and $\Psi_{-\nk} = \Psi_\nk^*$ because $y(\x,\eta)$ and $\Psi(\x,\eta)$ are real. Substituting the Fourier expansions into Lagrangian \eqref{action2}, the resulting action is $\delta^{(2)} S = \int d\eta  {L}^{(2)}$, with ${L}^{(2)} \equiv \int_{\mathbb{R}^{3+}} d^3 k  \mathcal{L}^{(2)}$, 
\begin{eqnarray}\label{lagrangianok}
 \mathcal{L}^{(2)} &\equiv&   y_\nk' y_\nk^{*'} - (  k^2 - \frac{a''}{a} + a^2   V_{, \phi \phi}  ) y_\nk y_\nk^*  \nn
 & +& 4 a \phi_0' ( \Psi'_\nk y_\nk^* +  \Psi^{*'}_\nk y_\nk)  -2 a^3    V_{,  \phi} ( \Psi_\nk y_\nk^* + \Psi_\nk^* y_\nk). \nn
\end{eqnarray}
Note that we are defining ${L}^{(2)}$ by integrating the function  $\mathcal{L}^{(2)}$ over the $\nk^+$ half-space.

The CSL model is based on a non-unitary modification to the Schr\"odinger equation; consequently, it will be advantageous to perform the quantization of the perturbations in the Schr\"odinger picture, where the relevant physical objects are the Hamiltonian and the
wave functional. 

We first define the canonical conjugated momentum associated to $y_{\nk}$ is $p_{\nk} \equiv
\partial \mathcal{L}^{(2)}/ \partial y_{\nk}^{\star '}$, that is $p_{\nk} =  y_{\nk}'$. The Hamiltonian associated to Lagrangian ${L}^{(2)}$, can be found as $H^{(2)} =  \int_{\mathbb{R}^{3+}} d^3k \: (y^{*'}_\nk p_\nk + y^{'}_\nk p^*_\nk   ) -  {L}^{(2)}$. Therefore, ${H}^{(2)} = \int_{\mathbb{R}^{3+}} d^3 k \mathcal{H}^{(2)}$, with
\begin{eqnarray}\label{Hamiltclas}
\mH^{(2)} &\equiv&    p^*_\nk p_\nk  + y^*_\nk y_\nk \left( k^2 - \frac{a''}{a}  + a^2  V_{, \phi \phi}    \right)  \nn
& -& 4 a \phi_0' ( \Psi'_\nk y_\nk^* +  \Psi^{*'}_\nk y_\nk)  + 2 a^3    V_{,  \phi} ( \Psi_\nk y_\nk^* + \Psi_\nk^* y_\nk). \nn  
\end{eqnarray}

From the Hamiltonian above we can find the equation of motion for $y_\nk$ and $p_\nk$. That is, using that
\begin{equation}
p'_\nk = - \frac{\partial \mH^{(2)} }{\partial y^*_\nk }, \qquad y^{*'}_\nk =  \frac{\partial \mH^{(2)} }{\partial p_\nk }, 
\end{equation}
the field's mode equation of motion is
\begin{equation}
y_\nk'' + \left( k^2 - \frac{a''}{a}  +  a^2 V_{, \phi \phi}        \right)y_\nk   - 4 a \phi_0'  \Psi'_\nk   + 2 a^3    V_{,  \phi}  \Psi_\nk = 0.
\end{equation}
The previous equation coincides with the evolution equation for $\dphi_\nk$ usually found in the literature \cite{mukhanovbook}, thus it serves as a self-consistency check.

Given that we are carrying out the quantization in the Schr\"odinger picture, it will be more convenient to work with real variables, which later can be associated to Hermitian operators. Therefore, we introduce the following definitions
\begin{equation}
y_\nk \equiv \frac{1}{\sqrt{2}} (y_\nk^\text{R} + i y_\nk^\text{I}  ), \qquad p_\nk \equiv \frac{1}{\sqrt{2}} (p_\nk^\text{R} + i p_\nk^\text{I}  ),
\end{equation}
and also
\begin{equation}
\Psi_\nk \equiv \frac{1}{\sqrt{2}} (\Psi_\nk^\text{R} + i \Psi_\nk^\text{I}).
\end{equation}

In the Schr\"odinger approach, the quantum state of the system is described by a wave functional, $\Phi[y(\x,\eta)]$. In Fourier space (and since the theory is still free in the sense that it does not contain terms with power higher than two in the Lagrangian), the wave functional can also be factorized into mode components as
\begin{equation}
\Phi[y(\x,\eta)] = \prod_{\nk}  \Phi_\nk ( y_\nk^\text{R} , y_\nk^\text{I} ) =  \prod_{\nk}  \Phi_\nk^\text{R} ( y_\nk^\text{R} )  \Phi_\nk^\text{I} ( y_\nk^\text{I} ).
\end{equation}

Quantization is achieved by promoting $y_\nk$ and $p_\nk$ to quantum operators,  $\hat y_\nk$ and $\hat p_\nk$, and by requiring the canonical commutation relations,
\begin{equation}
[\hat y_\nk^\text{R,I}  , \hat p_\nq^\text{R,I}  ] = i \delta(\nk - \nq).
\end{equation} 
In the field representation, the operators would take the form:
\begin{equation}
\hat y_\nk^\RI \Phi =  y_\nk^\RI \Phi, \qquad \hat p_\nk^\RI \Phi = -i \frac{\partial \Phi }{\partial y_\nk^\RI}.
\end{equation}

For the moment let us put aside the CSL mechanism, and analyze the standard evolution of the wave function. The wave functional $\Phi$ obeys the Schr\"odinger equation which, in this context, is a functional differential equation. However, since each mode evolves independently, this functional differential equation can be reduced to an infinite number of differential equations for each $\Phi_\nk$. Concretely, we have
\begin{equation}
i \frac{\partial \Phi_\nk^\RI }{ \partial \eta} = \hat H^\RI_\nk  \Phi_\nk^\RI,
\end{equation}
where the Hamiltonian densities $\hat H^\RI_\nk $, are related to the Hamiltonian as $\hat{H}^{(2)} = \int_{\mathbb{R}^{3+}} d^3 k (\hat H^\text{R}_\nk  + \hat H^\text{I}_\nk ) $, with the following definitions
\begin{eqnarray}\label{HamiltRI}
\hat H^{R,I}_\mathbf{k } &=& \frac{(\hat p_\mathbf{k }^{R,I} )^2  }{2 }+ \frac{(\hat y_\mathbf{k }^{R,I} )^2  }{2 }  \left( k^2 - \frac{a''}{a}  + a^2  V_{, \phi \phi}    \right)   \nn
& -& 4 a \phi_0'  \Psi_\nk^{'\RI} \hat y_\nk^\RI  + 2 a^3    V_{,  \phi} \Psi_\nk^\RI \hat y_\nk^\RI. 
\end{eqnarray}

The standard assumption is that, at an early conformal time $\tau \to -\infty$, the modes
are in their adiabatic ground state, which is a Gaussian centered at zero with certain
spread. In addition, this ground state is commonly referred to as the Bunch-Davies (BD) vacuum. Thus, the conformal time
$\eta$ is in the range $[\tau,0^{-})$.

Since the initial quantum state is  Gaussian and the Hamiltonian (as well as the collapse Hamiltonian, see Eq. \eqref{HCSL}) is quadratic in $\hat y_\nk^\text{R,I}$ and $  \hat p_\nk^\text{R,I}$, the form of the state vector in the field basis at any time is
\begin{equation}\label{psionday}
\Phi^{R,I}(\eta,y_{\nk}^{R,I}) = \exp[- A_{k}(\eta)(y_{\nk}^{R,I})^2 +
B_{k}(\eta)y_{\nk}^{R,I} +  C_{k}(\eta)].
\end{equation}

Therefore, the wave functional evolves according to Schr\"odinger equation, with initial 
conditions given by
\begin{equation}
A_k (\tau ) = \frac{k}{2}, \qquad B_k (\tau ) = C_k(\tau )= 0.
\end{equation}

Those initial conditions correspond  to the BD vacuum, which is perfectly 
homogeneous and isotropic in the sense of a vacuum state in quantum field theory. 


After the identification of the Hamiltonian that results in Schr\"odinger's equation, which from now on we refer to as the ``free Hamiltonian,'' we now incorporate the CSL collapse mechanism.

The main physical idea  of the CSL model is that an objective reduction of the wave function occurs all the time for all kind of particles. The reduction or \textit{collapse} is spontaneous and random. The collapse occurs whether the particles are isolated or interacting and whether the particles constitute a macroscopic, mesoscopic or microscopic system.

In order to apply the CSL model to the inflationary setting, we will consider a particular version of the CSL model in which the nonlinear aspects of the CSL model are shifted to the probability law. Specifically, the evolution equation is linear, which is similar to Schr\"odinger's equation; however, the law of probability for the realization of a specific random function, becomes dependent of the state that results from such evolution. In other words,  the theory can be characterized in terms of two equations: 

The \textit{first} is a modified Schr\"odinger equation, whose solution is
\begin{equation}\label{CSLQM}
|\psi,t\rangle={\cal T}e^{-\int_{0}^{t}dt'\big[i\hat H + \frac{1}{4\lambda}[w(t')-2\lambda \hat A]^{2}\big]}|\psi,0\rangle.
\end{equation}
$\cal T$ is the time-ordering operator. The modified Schr\"odinger's equation given by \eqref{CSLQM}, induces the collapse of the wave function towards one of the possible eigenstates of $\hat A$, which is called the \textit{collapse generating operator}.  The parameter $\lambda$ is the universal CSL parameter that sets the strength  of the collapse. In particular, $\lambda$ serves to characterize the rate at which the wave function increases its ``localizations'' in the eigen-basis of the collapse operator. In laboratory situations, the collapse operator is usually chosen to be the position operator and $\lambda$ is assumed proportional to the mass of the particle \cite{bassi,bassi2012}; in this model, the collapse rate, which has dimensions of [Time]$^{-1}$, is given by $\lambda a^2$ (here $a$ is a second parameter that sets the correlation length above which spatial superpositions are reduced).

The function $w(t)$ describes a stochastic process (i.e. is a random classical function of time) of white noise type. In other words, CSL regards the state vector undergoing some kind of Brownian motion. The probability for  $w(t)$ is given by the \textit{second equation}, the Probability Rule
\begin{equation}\label{probCSL}
P[w(t)] Dw(t) \equiv \langle\psi,t|\psi,t\rangle\prod_{t_{i}=0}^{t}\frac{dw(t_{i})}{\sqrt{ 2\pi\lambda/dt}}.
\end{equation}
The norm of $ \bra \psi,t | \psi,t \ket $ evolves dynamically, i.e. does not equal 1. Hence, Eq. \eqref{probCSL} implies that state vectors with largest norm are most probable. Furthermore, the total probability satisfies $\int P D w(t) = 1$. The stochastic term also prevents a wave-packet from spreading indefinitely, and causes the width of the packet to reach a finite asymptotic value \cite{bassi2012}.

In the case of multiple identical particles in three dimensions, the CSL theory would contain one stochastic function for each independent degree of freedom $w^i (t)$, but only
one parameter $\lambda$. In the case of several species of particles,
the theory would naturally involve a parameter $\lambda_i$ for each
particle species. In fact, there is strong
phenomenological preference for a $\lambda_i$ that depends on the
particle's mass $m_i$ \cite{grw,per}.

Returning to the inflationary context, in Ref. \cite{pedrocsl} it is shown that with the appropriate selection of the field collapse operators and using the corresponding CSL evolution law one obtains collapse in the relevant operators corresponding to the Fourier components of the field and the momentum conjugate of the field.\footnote{We also acknowledge at this point that there is no complete version of the CSL theory that is applicable universally, ranging from the laboratory setting to the cosmological one. However, we adopt the point of view that proposing educated guesses, in combination with phenomenological models applicable to particular situations, allow us to advance in our program. } We further assume that the reduction mechanism acts on each mode of the field independently. Also, it is suitable to choose $\hat y_\nk$ as the collapse operator because our main equation \eqref{masterpsi} suggests that $\bra \hat y_\nk \ket$ can act as a source of the Newtonian potential.   Therefore, the evolution of the state vector  characterizing the inflaton  as given by the CSL theory  is assumed to be:
\begin{eqnarray}\label{CSLevolution}
|\Phi_{\nk}^{\textrm{R,I}}, \eta \ket &=& \hat T \exp \bigg\{ - \int_{\tau}^{\eta} 
d\eta'   \bigg[ i \hat{H}_{\nk}^{\textrm{R,I}} \nn
&+& \frac{1}{4 \lambda_k} (\mathcal{W}(\nk,\eta') - 2 \lambda_k
\hat{y}_{\nk}^{\textrm{R,I}})^2 \bigg] \bigg\} |\Phi_{\nk}^{\textrm{R,I}}, \tau \ket, \nn
\end{eqnarray} 
$\hat T$ is the time-ordering operator, and recall that $\tau$ denotes the conformal time at the beginning of inflation.


The parameter $\lambda_k$ is a phenomenological generalization of the CSL parameter, and now it depends on the mode $k$. From the point of view of pure dimensional analysis, $\lambda_k$ must have dimensions of [Length]$^{-2}$. Hence, the simplest parameterization we can assume is $\lambda_k = \lambda_0 k$, which is also the same parameterization considered e.g. in \cite{pedrocsl,pia_tesis}. At first glance, one can postulate that $\lambda_0$ should coincide with the empirical bounds obtained from laboratory experiments for the collapse rate. For example, the value $\lambda_{\textrm{GRW}} = 10^{-16}$s$^{-1}$ was originally proposed by Ghirardi, Rimini and Weber \cite{grw} and later adopted by Pearle \cite{per}  for his CSL theory, as providing sound behavior when applied to laboratory contexts. However, as argued in \cite{commentshort} there is no particular reason why one should expect that the collapse rate associated to the parameter $\lambda$, utilized in applications of the CSL model at present day laboratory situations (and whose values are probably tied to underlying atomic structure that did not exist in inflationary times), should necessarily, or even naturally, be the ones utilized in modeling the inflationary regime, i.e. $\lambda_0$. In Sec. \ref{secPSanalysis}, we will say more about $\lambda_k$ and the particular value(s) of $\lambda_0$ considered in the present work.


Additionally, we postulate that the white noise $w(t)$,  which appears in Eqs. \eqref{CSLQM} \eqref{probCSL}, is now a stochastic field that depends on $\nk$ and the conformal time. That is, since we are applying the CSL collapse dynamics to each mode of the field, it is natural to introduce a stochastic function for each independent degree of freedom. Henceforth, the stochastic field $\mathcal{W}(\nk,\eta)$ might be regarded as a Fourier transform on a stochastic spacetime field $\mathcal{W}(\vec x,\eta)$. Here we would like to mention that the generalization from $w(t)$  to $w(\x,t)$ is in fact considered in standard treatments of non-relativistic CSL models \cite{pearlemisc}. For instance, the generalization of the CSL model of a single particle in one dimension to a single particle in three dimensions, implies to consider  a joint basis of operators $\hat A^i$, which commute $[\hat A^i, \hat A^j] =0$, instead of single collapse operator $\hat A$. This change requires one white noise function $w^i(t)$ for each $\hat A^i$. A further generalization is to consider a ``continuum'' collapse operator $\hat A(\x)$ (e.g. the mass density operator smeared over a spherical volume),  this requires a random noise field $w(\x,t)$ instead of a set of random functions \cite{bassi,bassi2012,per,pearle3}.

Continuing with the calculations, we can take the time derivative of \eqref{CSLevolution} (see \cite{per}), obtaining
\begin{equation}\label{Htotal}
\frac{\partial}{\partial \eta}|\Phi_{\nk}^{\textrm{R,I}}, \eta \ket = -    i \hat{H}_{\nk}^{\textrm{R,I}} 
+ \hat{H}_{\nk\: \textrm{CSL}}^{\textrm{R,I}}  |\Phi_{\nk}^{\textrm{R,I}}, \eta \ket,
\end{equation}
with
\begin{equation}\label{HCSL}
\hat{H}_{\nk\: \textrm{CSL}}^{\textrm{R,I}} \equiv - \frac{\mathcal{W}(\nk,\eta)^2}{4 \lambda _k} + \mathcal{W}(\nk,\eta)  \hat{y}_{\nk}^{\textrm{R,I}} -  (\hat{y}_{\nk}^{\textrm{R,I}})^2 \lambda_k.
\end{equation}

Next, taking into account that our main goal is to obtain the primordial power spectrum, see Eq. \eqref{masterPS},  we turn our attention to compute the quantity $\overline{\bra \hat y_\nk \ket \bra \hat y_\nq \ket^* }$. The expectation values of course will be evaluated at the evolved state provided by \eqref{CSLevolution}. 

In terms of the real and imaginary parts, we have
\begin{equation}\label{masterpromedios}
\overline{\bra  \hat y_\nk \ket \bra \hat y_\nq \ket^* } =  \left( \overline{ \bra \hat  y_\nk^\text{R} \ket^2  }  + \overline{ \bra \hat  y_\nk^\text{I} \ket^2  }         \right) \delta(\nk-\nq ).
\end{equation}
Note that we have assumed that the CSL model does not induce modes correlations. Also from \eqref{masterpromedios}, it is clear that we are interested in computing the quantities $\overline{ \bra \hat  y_\nk^\text{R,I} \ket^2  } $. In fact, the calculation of the real and imaginary part are exactly the same, so we will only focus on one of them. Additionally, 
we simplify the notation by omitting the indexes R,I from now on. 

Using the Gaussian wave function \eqref{psionday}, and the CSL evolution equations, it can be shown \cite{pedrocsl} that 
\begin{equation}\label{masterresta}
\overline{ \bra \hat  y_\nk \ket^2  } = \overline{ \bra \hat  y_\nk^2 \ket  } - \frac{1}{4 \textrm{Re}[A_k (\eta)]  }.
\end{equation}
Therefore, in order to obtain a prediction for the power spectrum, we need to calculate the two terms on the right hand side of \eqref{masterresta}. Explicit computation of Eq. \eqref{masterresta} implies solving the corresponding CSL equations. At this point we would like to mention that the actual calculations are long and cumbersome, but we have include them in Appendix \ref{csl_eq_appendix} for the interest reader. The final result corresponding to the quantity $\overline{ \bra \hat  y_\nk \ket^2  } $ is given in \eqref{masterycuadrado}.

Another final remark regarding Eq. \eqref{masterresta} is that the right hand side is technically easier to handle than to directly solve the main CSL evolution equation for the state vector \eqref{Htotal}. The latter procedure is difficult because the Hamiltonian $\hat H_\nk$ given by \eqref{HamiltRI}, contains terms that involve $\Psi_\nk'$ and $\Psi_\nk$. These terms in turn depend on the state vector through the expectation value $\bra \hat  y_\nk \ket$ as shown in the main equation \eqref{masterpsi}.  On the other hand, given that we are only interested in computing the power spectrum, the computation of the right side of \eqref{masterresta} allow us to bypass such a direct calculation. The details can be found in Appendix \ref{csl_eq_appendix}, but here we can mention, that for instance, the evolution equation for $A_k(\eta)$ decouples from the other quantities $B_k(\eta)$ and $C_k(\eta)$ which do involve a more elaborated method to solve. Moreover, the evolution equation for $\overline{ \bra \hat  y_\nk^2 \ket  }$  does not involve the linear term $ \bra \hat  y_\nk \ket$ only [see Eqs. \eqref{evolQRS} and Eq. \eqref{m2}].  

We acknowledge that this is a pragmatic way to proceed and that formally one would have to perform the full quantization using  Hamiltonian \eqref{Htotal}, which means the formal characterization of the collapse process within the semiclassical treatment. Some advances in this direction have been made \cite{alberto,erandy}. The main idea developed in those works is that any sudden change in the quantum state (i.e. a collapse) will result, generically, in a sudden modification in the expectation value of the energy-momentum tensor, and thus to a different space-time metric. Nonetheless, such modification would in general, require also a change in the quantum field theory construction, and consequently a new Hilbert space to which the state can belong. In this way, one would have a QFT and a space-time metric corresponding to the initial state, and a different QFT/metric for the post-collapse state. Then the two different space-times must be glued together in a consistent way. While this is the correct framework to adopt, it is beyond the scope of the present paper. Instead we will focus on computing the right hand side of \eqref{masterresta} using two reasonble assumptions: (i) the collapse generating operator is $\hat y_\nk$; therefore, we expect that the modified Schr\"odinger equation \eqref{CSLevolution} drive the initial Gaussian state to a final state that is very similar to an eigenvector of $\hat y_\nk$. That is to say, the final wave function can be approximated by a Dirac function $\delta(y_\nk - Y_\nk)$, where $Y_\nk \equiv \bra \hat y_\nk \ket$ evaluated in the final state. (ii) we will assume that the localization process is fast enough compared to the total duration of inflation, in conformal time this is $\lambda_0 |\tau| > 1$; so the time evolution of $\Psi(\x,\eta)$ is deterministic [in fact given by Eq. \eqref{psievo}]. In Appendix \ref{csl_eq_appendix} we present the computational details of Eq. \eqref{masterresta}  using those two assumptions [in particular to obtain Eqs. \eqref{evolQRS2} and \eqref{evolAk}].


\section{Analysis of the primordial and angular power spectra}\label{secPSanalysis}

Given the solutions of the CSL equations, we can obtain the power spectrum. Clearly, this allow us to compare the predictions between the standard model and our proposal. 

The path is straightforward:  we substitute $\overline{ \bra \hat  y_\nk \ket^2  } $ [whose explicit form is shown in \eqref{masterycuadrado}] into \eqref{masterPS}, this yields the power spectrum. The detailed calculations can be found in Appendix \ref{app:calculationPs}, and the resulting expression of  $\mP_s(k)$ is given in \eqref{PSdiamond}. Such an expression represents the primordial power spectrum at  second order in the HFF, and can also be  used to obtain $n_s$ and $\alpha_s$ at third and fourth order in the HFF respectively  [see Appendix \ref{app:calculationPs}, Eqs. \eqref{nscsl} and \eqref{alphacsl}]. 

Using expressions for $\mP_{s}$, $n_s$ and $\alpha_s$ at second order in the HFF allow us to parameterize the primordial power spectrum in terms of the scalar spectral index and its running, this is
\begin{equation}\label{pdek}
\mP_s(k) = A_s \left( \frac{k}{k_\di} \right)^{n_s-1 + \frac{\alpha_s}{2} \ln \frac{k}{k_\di}       } C(k)
\end{equation}
with
\begin{equation}
A_s = \frac{ H_\di^2}{ \pi^2 M_P^2 \epsilon_{1\di}},
\end{equation}
and the function $C(k)$ expressed in terms of $n_s$ (scalar spectral index) and $\alpha_s$ (running of the spectral index) is  
\begin{widetext}
\begin{eqnarray}\label{defCK2}
 C(k) &=&   1 + \frac{\lambda_k   |k \tau|}{ k^2} + \frac{\lambda_k }{2 k^2} \cos(2 |k \tau|)    \nn 
 &-&   \exp \bigg\{   \bigg[ -4+ n_s + \alpha_s  \ln \frac{2 k}{k_\di}    \bigg]  \ln \zeta_k  - \frac{\alpha_s}{2}  \left( \ln^2 \zeta_k - \theta_k^2  \right)   \bigg\}  \nn
 &\times& \bigg[ \cos \bigg\{ \left[  -4 +n_s + \alpha_s \ln \frac{2 k}{k_\di}  \right] \theta_k 
 - \alpha_s  \left(  -\Delta N_\di + \frac{2}{3} + D + \ln \frac{k}{2 k_\di}  \right) \theta_k \ln \zeta_k  \bigg\} \bigg]^{-1},
\end{eqnarray}

\end{widetext}
where $k_\diamond = 0.05$  Mpc$^{-1}$ is a pivot scale, $D \equiv 1/3 - \ln 3 $ and $\Delta N_\di$ is the number of e-folds from the horizon crossing of the pivot scale to the end of inflation, typically  $\Delta N_\di \sim 60$. The quantities $\theta_k$ and $\zeta_k$ are defined as:
\begin{equation}\label{defzetakythetak_main}
\zeta_k \equiv \left( 1 + \frac{4 \lambda_k^2}{k^4}   \right)^{1/4}, \qquad \theta_k \equiv - \frac{1}{2} \arctan \left( \frac{2 \lambda_k}{k^2} \right).
\end{equation}



We note that if $\lambda_k=0$, which means $\zeta_k=1$ and $\theta_k =0$, one can check that $C(k)=0$, hence $\mP(k)=0$. This is consistent with our model in which the collapse of the wave function, given by the CSL mechanism, is the source of the metric perturbations. Therefore, in our approach if there is no collapse, and the vacuum state remains unchanged there are no primordial perturbations, thus, $\mP(k)=0$ because $\Psi_{\nk} =0$ at all scales.

On the other hand, if  $\lambda_k/k^2 \gg 1$  then $\zeta_k \gg 1$ and $\theta_k \simeq -\pi/4$. This means that the evolution of the wave function is certainly being affected by the extra terms added by the CSL evolution equation. We recall that in the previous section we proposed the parameterization,
\begin{equation}\label{parametrizacion}
\lambda_k = \lambda_0 k,
\end{equation}
where $\lambda_0$ can be related to the universal CSL rate parameter, which has units of [Time]$^{-1}$. Therefore, $\lambda_0^{-1}$ provides us with a localization time scale for the wave function associated to each mode of the field. For the purpose of our analysis, we set the numerical value of the CSL parameter as  $\lambda_0 = 10^{-14}$ s$^{-1}$, or equivalently $1.029$ Mpc$^{-1}$ in the units chosen for the present work. This value is two orders of magnitude greater than the historical value $\lambda_{\textrm{GRW}} = 10^{-16}$ s$^{-1}$ suggested by GRW \cite{grw}, and also consistent with empirical constraints obtained from different experimental data such as:  spontaneous X-ray emission \cite{sandro2017}, matter-wave interferometry \cite{toros2016}, gravitational wave detectors \cite{carlesso2016} and  neutron stars \cite{cota_lambda0}. Also, according to assumption (ii) mentioned at the end of Sec. \ref{Sec3}, we have chosen the value $|\tau|= 7803894$ Mpc, so $\lambda_0 |\tau| = 8 \times 10^{6} > 1$. 
Our proposed parameterization  in Eq. \eqref{parametrizacion} is the most simple one for a $k$ dependence in the 
$\lambda_k$ parameter, although it is not the only possibility. As we shall see in the following analysis, the CSLIM induces oscillatory 
features in  $\mP_s (k)$ that  remain at some scales, showing that the effect of the collapse cannot be ``turned off''. This can be explained by noting first that, since the modes $k$ are infinite, there will be some modes $k$ such that the condition $\lambda_k/k^2 \gg 1$ fails. However, for the chosen value $\lambda_0 = 1.029$ Mpc$^{-1}$ and at least for the range of modes of observational interest $10^{-6}$ Mpc$^{-1}$ $ \leq k \leq 10^{-1}$ Mpc$^{-1}$, i.e. the ones that contribute the most to the CMB angular spectrum, the condition $\lambda_k/k^2 = \lambda_0/k \gg 1$ is fulfilled.  It is for these range of $k$ that we will analyze the features in  $\mP_s (k)$ induced by the CSLIM.

\begin{figure}[h]
  \centering
  \includegraphics[scale=0.35,angle=270]{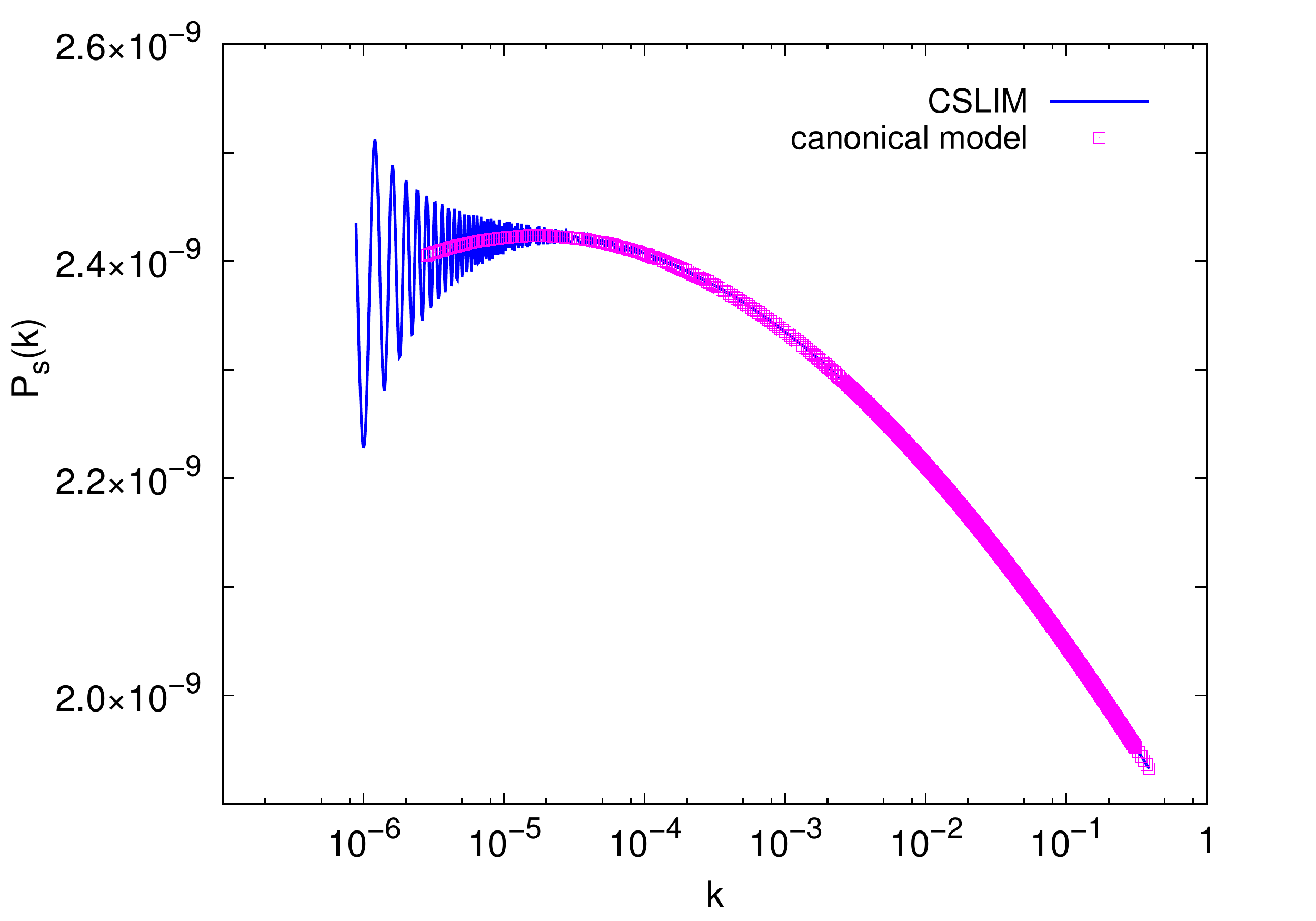}
  \caption{Comparison between the CSLIM power spectrum and the canonical model. The wave number $k$ is given in Mpc$^{-1}$. A good agreement is shown at high scales, while for $k < 0.0001$ Mpc$^{-1}$ oscillatory features introduced by the CSLIM become evident. }
  \label{fig:Pk_oscilaciones}
\end{figure}

In order to analyze those novel features, we plot expression \eqref{pdek}. The resulting plot is shown in Fig. \ref{fig:Pk_oscilaciones}, together with the prediction corresponding to the standard model, the latter being essentially Eq. \eqref{pdek} with $C(k) = 1$. The values of the inflationary parameters we have used are: $n_s= 0.9641$ and $\alpha_s = -0.0045$. Oscillations appear at low scales while no difference at all can be found for $k > 0.0001$ Mpc$^{-1}$. In fact the oscillations, induced by the CSLIM around the standard spectrum, show a decrease in amplitude at the higher end of scales.





The next step in our analysis is to investigate whether the oscillations shown in the primordial power spectrum have any incidence in the observational predictions. However, we want to stress that, in this paper, we will only perform a preliminary analysis of the CMB angular power spectrum (also known as the $C_l$ in the literature \cite{Planck2018_cosmo}) predicted by the CSLIM taking into account our second order power spectrum. A complete data analysis, including statistical analysis, is left for future work. Furthermore, we will limit ourselves to the analysis of the temperature auto-correlation spectrum; however, from a previous analysis of similar models \cite{pia_tesis} we might expect that the $E$-mode polarization and Temperature-$E$-mode cross correlation will also be modified as a consequence of the collapse hypothesis. 

In order to perform our analysis, we have modified the Code for Anisotropies in the
Microwave Background (CAMB) \cite{camb} as to include the CSLIM predictions, which only affect the inflationary part of the $\Lambda$CDM canonical model. The rest of the cosmological parameters remain unchanged. Let us define the cosmological parameters of the canonical model: baryon density in units of the critical density $\Omega_B h^2 = 0.02237$, dark matter density in units of the critical density $\Omega_{CDM} h^2 = 0.12$, Hubble constant in units of Mpc$^{-1}$ km s$^{-1}$ $H_0 = 67.36$. 
 We also include in that set the aforementioned values of $n_s$, $\alpha_s$ and $k$-pivot; all represent the best-fit values presented by the Planck collaboration \cite{Planck2018_cosmo}. The value of $A_s$ is $2.1 \times 10^{-9}$ for both CSLIM and canonical model.

\begin{figure}[h]
  \centering
  \includegraphics[scale=0.35,angle=270]{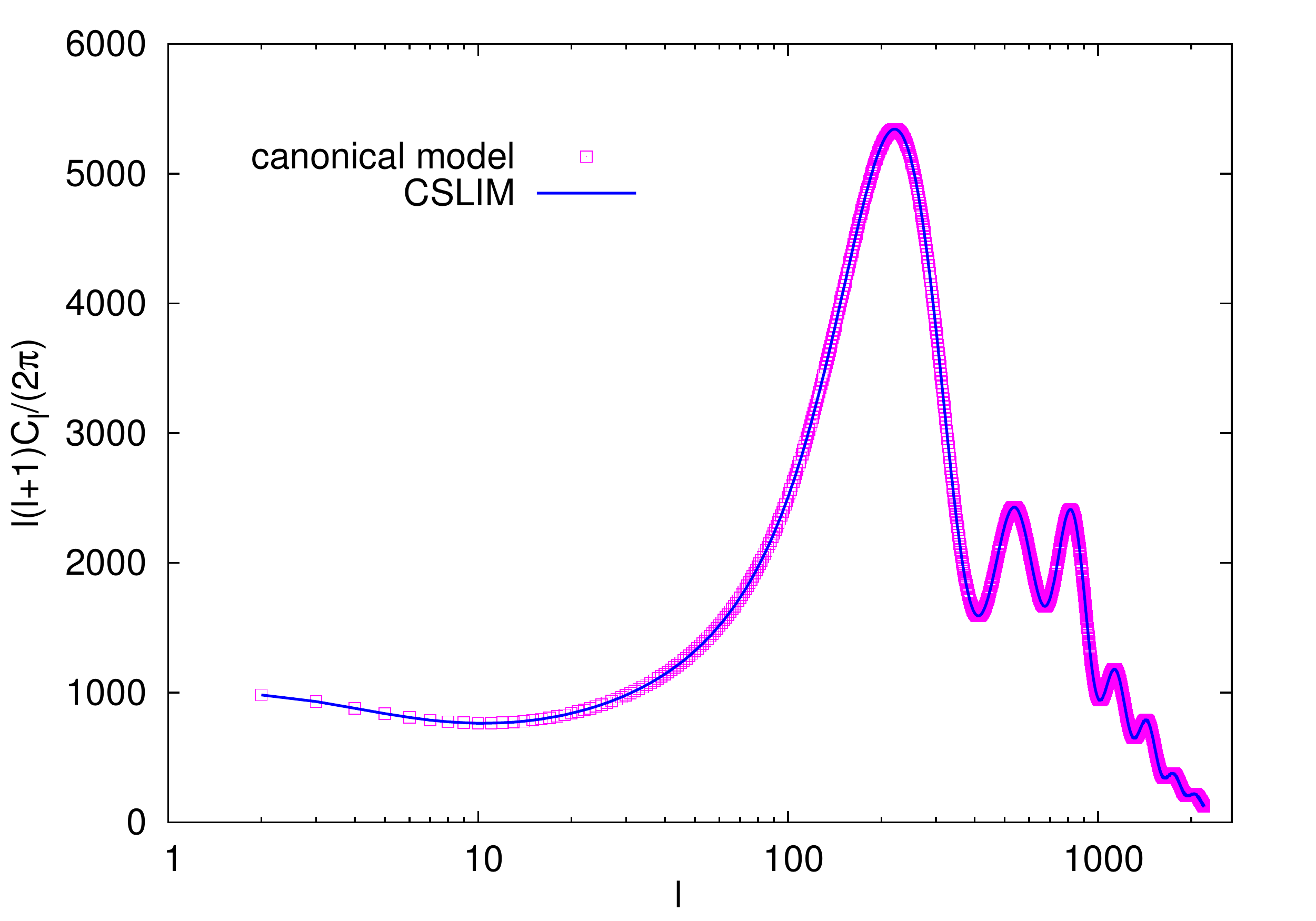}
  \caption{Temperature auto-correlation (TT) spectrum comparison between the canonical model (boxes) and the CSLIM (blue solid line), the latter using the parameterization $\lambda_k = \lambda_0 k$.  No difference is shown among them. Oscillations at low values of $k$ in the primordial power spectrum, as shown in  Fig. \ref{fig:Pk_oscilaciones}, are wiped off in the TT spectrum.}
  \label{fig:Cls}
  \end{figure}

Figure \ref{fig:Cls} shows the temperature auto-correlation (TT) spectrum for both CSL and canonical models, showing no
difference between them. As can be seen there, oscillating features at low $k$ in our predicted power spectrum
do not translate into any peculiar features in the theoretical predictions of the $C_l$ coefficients characterizing the angular temperature anisotropies. In this way, parameterization \eqref{parametrizacion} represents a good choice to set a basis for comparison with the canonical model, and in a sense also serves as a consistency check.

At this point of the analysis we have learned that $\lambda_k = \lambda_0 k$ yields an indistinguishable prediction from the canonical model. However, there is no reason to expect an exact $k$ dependence of $\lambda_k$. As a consequence, we proceed to explore possible effects of the CSLIM that can be reflected in the observational data by introducing a  new parameter $B$ through the parameterization of $\lambda_k$. The  role of $B$ will be to imprint a slight departure from the
canonical model shape. The new proposal to parameterize $\lambda_k$ is
\begin{equation}\label{parametrizacionB}
\lambda_k = \lambda_0 (k+B),
\end{equation}
 where $B$ has units of $\rm Mpc^{-1}$ and conforms a new parameter of the model that needs to be estimated
with recent observational data, this will be left for future work. In the rest of the present section, we will be interested in analyzing the consequences of varying $B$ on the predicted spectrum.

\begin{figure}[h]
  \centering
  \includegraphics[scale=0.37,angle=270]{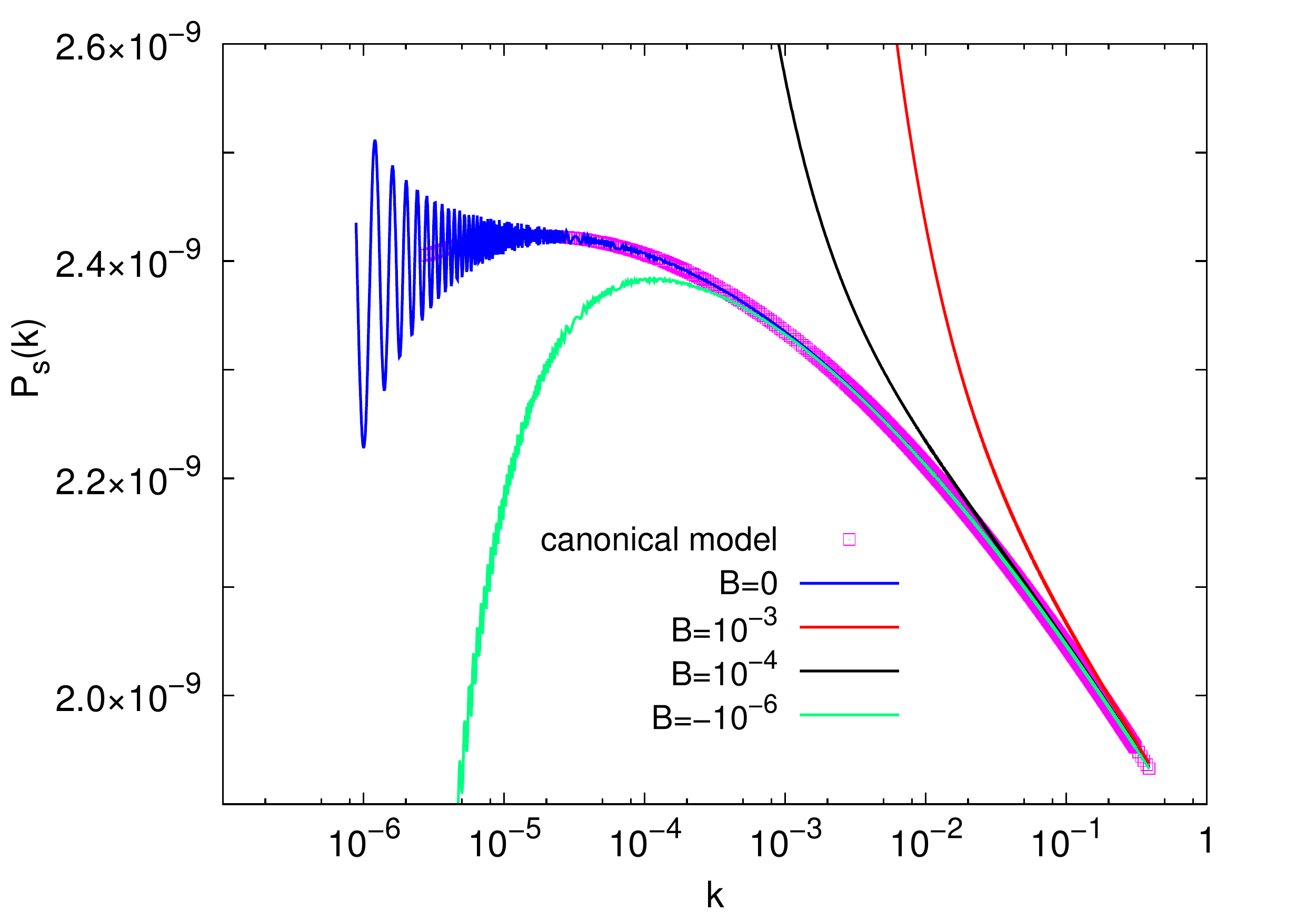}
  \caption{Here we appreciate small departures from the canonical power spectrum. The departures are parameterized by  $B$.  The set of $B$ values considered show effects properly attributed  to the CSLIM, and become explicitly manifest in the primordial power spectrum. Power suppression is seen at low scales for negative values of $B$, whereas positive values imply an upper departure from the canonical model. In this figure, $B$ and $k$ are given in  Mpc$^{-1}$.   }
  \label{fig:Ps_B}
\end{figure}

The effect of considering different values of $B$ on the power spectrum is shown in Fig. \ref{fig:Ps_B}, where the same plot of Fig. \ref{fig:Pk_oscilaciones} has been included as the case $B=0$, and serves as a reference. For negative $B$ (green line) the CSLIM power spectrum seems to approach to the canonical one from below, showing significant differences for $k<10^{-4}$ Mpc$^{-1}$. Meanwhile, for positive $B$ the CSLIM power spectrum approaches from above (black and red lines). The differences in the predicted spectrum between the CSLIM and the canonical seem to dissolve progressively as $B$ approaches zero, remaining only a small differences at low $k$ due to the oscillations. Also, it is worthwhile to mention that oscillations present in the $B=0$ case cannot be significantly appreciated in  the rest of cases.  Figure \ref{fig:Ps_B} suggests that observational predictions in the CSLIM may be distinguished from the ones of the canonical model. In the next final part, we analyze whether these departures (from the canonical model) have observable consequences on the CMB fluctuation spectrum.

\begin{figure}[h]
  \centering
  \includegraphics[scale=0.35,angle=270]{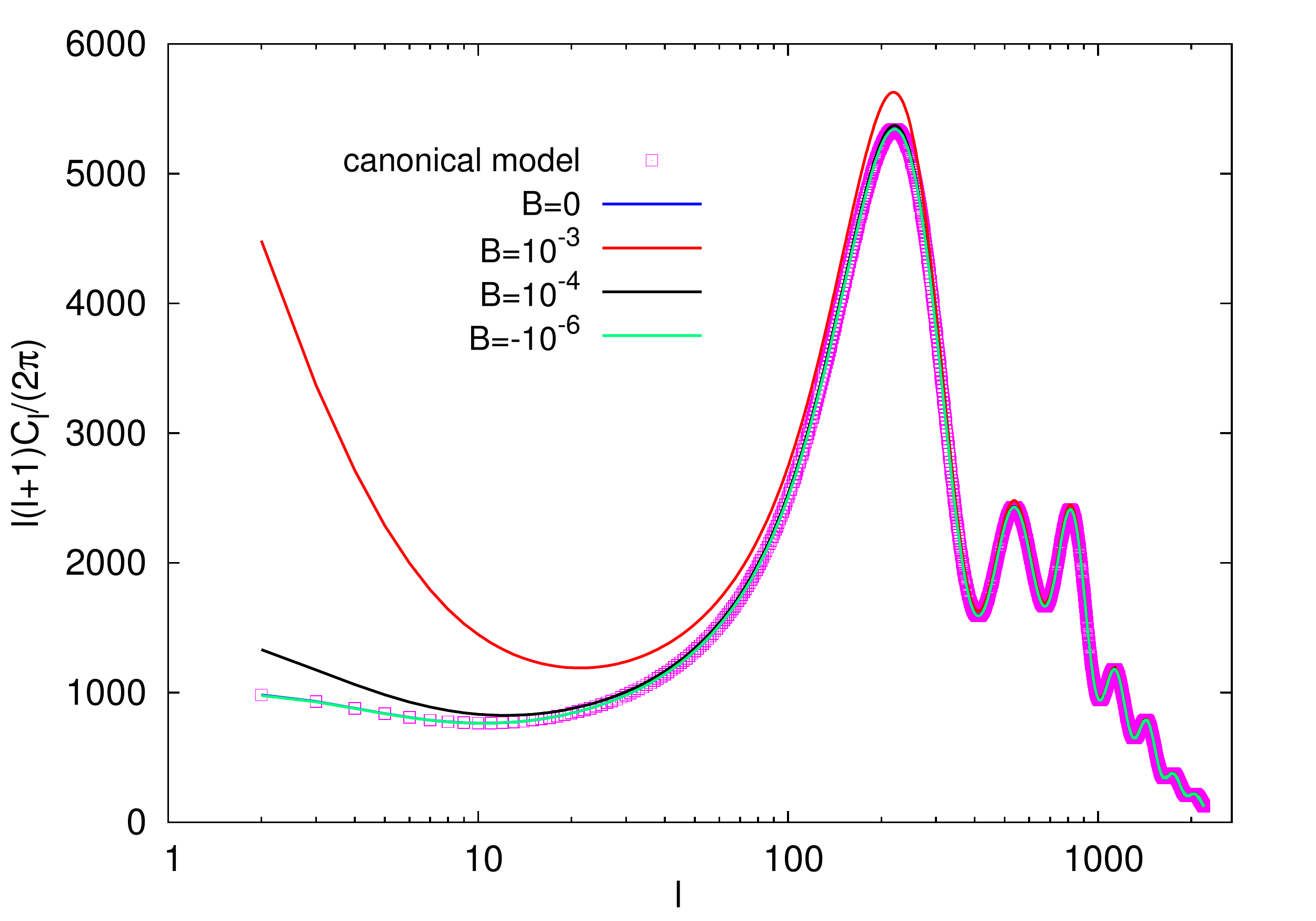}
  \caption{The temperature auto-correlation (TT) spectrum including the $B$ parameter, the values of the cosmological parameters considered are the same as in Fig. \ref{fig:Cls}. The negative $B$ value does not exhibit any difference at all with respect to the canonical model (boxes). On the other hand, for positive $B$ values there is a progressive departure from the canonical model at low $l$ (big angular scales). The case  $B=10^{-3}$ Mpc$^{-1}$ can be discarded in advance as it modifies substantially the position of the first peak, which is constrained at a high degree of accuracy by current data.   
}
  \label{fig:Cls_B}
\end{figure}

  Figure  \ref{fig:Cls_B} shows our prediction for the CMB temperature fluctuation spectrum and the canonical one, where we used the same values for the cosmological parameters as before. From Fig. \ref{fig:Cls_B}, it can be inferred an estimated upper limit for the $B$ parameter, i.e. for $B=10^{-3}$ Mpc$^{-1}$ the first peak is shifted upwards which is totally incompatible with the latest high precision observational measurements. The negative $B$ value tested does not induce any significant difference in parameter estimation when compared with the canonical model. In the case of $B=10^{-4}$ Mpc$^{-1}$, a small departure from the canonical prediction is seen at low multipoles. Whether this change is favored by the data or simply lost in the cosmic variance uncertainty will be addressed in future research. Nonetheless, from this analysis we can infer that in order for our predicted power spectrum to be consistent with the best fit temperature auto-correlation spectrum, and at the same time, to manifest departures from the canonical model, the $B$ parameter must be then constrained between $B>0$ and $B < 10^{-3}$ Mpc$^{-1}$. 
  

\subsection{Consequences of varying $\lambda_0$}

In this final part of the present section, we would like to make some remarks about how our previous results would be affected by considering different values of $\lambda_0$. As we have argued, $\lambda_0^{-1}$ may be used to set a localization time scale for the wave function associated to each mode of the field, so varying $\lambda_0$ means to change the localization time scale. 

The criteria to select appropriate values of $\lambda_0$ is based on the condition $\lambda_k/k^2 \gg 1$. If that condition is satisfied then the collapse occurs successfully; particularly, one would require that said condition is met for all  modes $k$ within the range of interest $10^{-6}$ Mpc$^{-1}$ $ \leq k \leq 10^{-1}$ Mpc$^{-1}$.  Moreover, we consider the parameterization $\lambda_k = \lambda_0 (k + B) $ constrained within the range $ 0 \leq B < 10^{-3}$ Mpc$^{-1}$. Therefore, we  note first that the condition $\lambda_k/k^2 = \lambda_0 (1 + B/k)/k \gg 1$  is fulfilled for the previous chosen value $\lambda_0 = 1.029$ Mpc$^{-1}$ (with $B$ and $k$ in the aforementioned ranges). Second, it is clear that in order to satisfy the condition $\lambda_k/k^2 \gg 1$ for different values of the parameter, we must consider  $\lambda_0 \geq 1.029$ Mpc$^{-1}$. 

We have reproduced Figs. \ref{fig:Pk_oscilaciones}, \ref{fig:Cls}, \ref{fig:Ps_B} and \ref{fig:Cls_B} for the values $\lambda_0 = 10.29$ Mpc$^{-1}$ and $\lambda_0 = 102.9$ Mpc$^{-1}$ obtaining exactly the same  plots as the ones corresponding to the original value $\lambda_0 = 1.029$ Mpc$^{-1}$. In particular, the shape of the spectra is exactly equal. However, to achieve a similar amplitude, we had to adjust the combination $V |\tau | /M_P^4\epsilon_1  $, we remind the reader that $\tau$ is the conformal time at which inflation begins.

In order to attain a better understanding of this result, we focus on our prediction for $\mP_{s} (k)$, Eq. \eqref{pdek}. To make things simple and without loss of generality, we assume $n_s =1$, $\alpha_s =0$ and we use Friedmann's equation $H^2 \simeq V/3 M_P^2$.  Consequently with these assumptions Eq. \eqref{pdek} is approximately
\begin{eqnarray}\label{Pkapprox}
\mP_s(k)  &\simeq& \frac{ V }{ 3 \pi^2 M_P^4 \epsilon_1}   \bigg( 1 + \frac{\lambda_k   |k \tau|}{ k^2} + \frac{\lambda_k }{2 k^2} \cos(2 |k \tau|)  \nn
&-& \frac{1}{\zeta_k^3 \cos 3 \theta_k}   \bigg).
\end{eqnarray}
From definitions \eqref{defzetakythetak_main},  $ \lambda_k/k^2 \gg 1$  implies that $\zeta_k \gg 1$ and $\theta_k \simeq -\pi/4$. Let us consider the parameterization $\lambda_k \simeq \lambda_0 k$; hence, if condition $ \lambda_k/k^2 \gg 1$ is met, then Eq. \eqref{Pkapprox} can be approximated by\footnote{In  approximation \eqref{Pkapprox2}, we also used the fact that  $k|\tau| > 1$. Recalling that $\tau$ is the conformal time at which inflation begins, $k |\tau| > 1$ is essentially satisfied by all the modes because such condition means that said  modes begun in the Bunch-Davies vacuum. }
\begin{equation}\label{Pkapprox2}
\mP_s(k) \simeq \frac{ V }{ 3 \pi^2 M_P^4 \epsilon_1} \lambda_0 |\tau |.
\end{equation}
Note that assumption (ii) mentioned at the end of Sec. \ref{Sec3} and Eq. \eqref{Pkapprox2} are consistent, this is  $\lambda_0 |\tau| > 1$. With result \eqref{Pkapprox2} at hand, we can now conclude that increasing $\lambda_0$ requires to decrease the combination $V |\tau | /M_P^4\epsilon_1  $, such that $\mP_s(k)$  would  be consistent with the observed amplitude $\mP_s(k) \simeq 10^{-9}$. For example, if one increases the value of $\lambda_0$ but $\epsilon_1$, $\tau$ remain fixed, then the characteristic energy scale of inflation $V^{1/4}$ must decrease. Also, note that using Eq. \eqref{Pkapprox} and the parameterization $\lambda_k \simeq \lambda_0 k$ lead to a scale invariant power spectrum (independent of $k$), Eq. \eqref{Pkapprox2}. This was expected since we considered $n_s =1$ and $\alpha_s=0$, but the main point is that varying $\lambda_0$ does not affect the shape of the spectrum. On the contrary, varying the parameterization of $\lambda_k$ would certainly alter the scale dependence of the spectrum, and  Figs. \ref{fig:Pk_oscilaciones}, \ref{fig:Cls}, \ref{fig:Ps_B} and \ref{fig:Cls_B} would also change substantially.


\section{Summary and Conclusions}
\label{Sec5}

In this work, we have calculated the primordial power spectrum for a single scalar field during  slow roll inflation. The calculation considered the application of the Continuous Spontaneous Localization (CSL) objective reduction model to the inflaton wave function, within the semiclassical gravity setting. The novel aspect in this paper was to consider the second order approximation in the Hubble flow functions (HFF), and solve the corresponding CSL equations using the uniform approximation method in slow roll inflation \cite{habib2004,venninSR}. 

The implementation of the CSL model to slow roll inflation or CSL inflationary model (CSLIM) for short, induced a modification of the standard scalar power spectrum (PS) of the form $\mP_s(k) = A_s k^{n_s-1 + \frac{\alpha_s}{2} \ln \frac{k}{k_\di} } C(k)$.  One of the main features uncovered here is that the function $C(k)$ depends on the inflationary parameters $n_s$, $\alpha_s$ as well as the collapse parameter $\lambda_k$, see Eq. \eqref{defCK2}.  

We have chosen the most simple parameterization for the collapse parameter, this is  $\lambda_k = \lambda_0(k + B)$, where $\lambda_0$ is the fundamental CSL parameter, representing the collapse rate, and $B$ is a new parameter. We have set $\lambda_0 = 10^{-14}$ s$^{-1}$ or $1.029$ Mpc$^{-1}$ (which is two orders of magnitude greater than the historical value suggested for the collapse rate \cite{grw}), and varied $B$ from $B = -10^{-6}$ Mpc$^{-1}$ to $B = +10^{-3}$ Mpc$^{-1}$; these values gave rise to  significative departures from the standard PS, see Fig. \ref{fig:Ps_B}, mostly at the lower range of $k$.  Next, we have shown the effects of the CSLIM on the CMB temperature fluctuation spectrum, Fig. \ref{fig:Cls_B}. For this preliminary analysis, the proposed parameterization of $\lambda_k$  seems to be in good agreement with the present data of the CMB fluctuation spectrum.  In particular, within the range $B = -10^{-6}$ Mpc$^{-1}$ and $B=0$ there are no differences between the prediction of the CSLIM and the standard inflationary model, in spite of the evident variations in the PS.
However, between $B>0$ and $B < 10^{-3}$ Mpc$^{-1}$ there are important departures from the standard model prediction in the temperature fluctuation spectrum but at the same time could be consistent with the best fit temperature auto-correlation spectrum. We have also shown that values $B \geq 10^{-3}$ Mpc$^{-1}$ could be discarded without performing any statistical analysis. Finally, we have argued that increasing $\lambda_0$ will not have an effect on the shape of the spectra but it can have theoretical consequences in the parameters characterizing the spectrum's amplitude, e.g. the characteristic energy scale of inflation.

Our result $\mP_s(k) = A_s k^{n_s-1 + \frac{\alpha_s}{2} \ln \frac{k}{k_\di} } C(k)$, with $C(k)$ depending explicitly on $\lambda_k$, $\alpha_s$ and $n_s$, allow us to identify exactly the dependence on $k$ attributed to: the CSL model, the spectral index and the running of the spectral index. We think this is an important result because of the following. Our predicted PS allows departures from the traditional inflationary approach that can be tested experimentally. As we have argued in the Introduction, if future experiments detect a significant value of the running of the spectral index, i.e. of order $|\alpha_s| \simeq 10^{-3}$ and the tensor-to-scalar ratio $r$ remains undetected,  then the hierarchy of the HFF would be broken and the standard slow roll inflationary model would be in some sense jeopardized. On the other hand, the CSLIM generically predicts a strong suppression of tensor modes, that is $r \simeq \epsilon_{1 \di}^2 10^{-12}$ \cite{nobmodesbig,nobmodesshort}. And, since the function $C(k)$ introduces an extra $k$ dependence on the PS, the situation described previously, in principle, could not yield an inconsistency between the CSLIM and hierarchy of the HFF. Specifically,  what in the standard approach might be identified as a running of the spectral index, which is essentially a particular dependence on $k$ of the PS,  in the CSLIM the same effect could be attributed to $C(k)$ through the parameterization of $\lambda_k$, and in particular to the $B$ parameter. In other words, in the CSLIM, the hierarchy $|\epsilon_{1 \di}| > |\epsilon_{2 \di}| > |\epsilon_{3 \di}|$ could be satisfied and still be consistent with observations, namely a non-detection of primordial gravity waves and a particular shape of the PS characterized by a ``running of the spectral index'' in the standard approach. 

Evidently, to test if the above conjecture is true, we require to perform a complete statistical analysis using the most recent (and future) observational data from the CMB. In particular, we would  be able to constrain the value of $B$ as well as $n_s$ and $\alpha_s$ within our model. Nevertheless, our main conclusion is that the CSLIM possess observational  consequences, different from the standard inflationary paradigm. In fact, some particular observations that would cause some issues in the traditional model,  could be potentially resolved within our approach. A final important lesson to be drawn from this analysis is that it displays how, at least in applications to cosmology, considerations regarding \textit{the quantum measurement problem} can lead to striking alterations concerning observational issues. This contributes to oppose a posture that claims such questions as of mere philosophical interest and dismisses their relevance regarding physical predictions.

%
%

\begin{acknowledgments}
G.L. and M.~P.~P are supported by CONICET (Argentina) and the National Agency for the Promotion of Science and Technology (ANPCYT) of Argentina grant PICT-2016-0081. M.~P.~P is also supported by grants G140 and G157 from Universidad Nacional de La Plata (UNLP).

\end{acknowledgments}

\appendix

\section{Solving the CSL equations}\label{csl_eq_appendix}

We begin by writing some useful expressions involving $V$, $\partial_\phi V$ and $\partial^2_{\phi \phi} V$ in terms of the Hubble flow functions (HFF). Therefore, one has the following quantities \cite{liddleSR,terreroSR}
\begin{equation}
\frac{M_P^2}{2} \left(  \frac{\partial_\phi V}{V}  \right)^2  = \epsilon_1 \left( 1 + \frac{\epsilon_2}{2(3-\epsilon_1)}     \right)^2,
\end{equation}
\begin{equation}\label{Vpp}
M_P^2  \frac{\partial^2_{\phi \phi} V}{V}  = \frac{6 \epsilon_1 - 3 \epsilon_2/2 -2 \epsilon_1^2 - \epsilon_2^2/4 + 5 \epsilon_1 \epsilon_2 /2 - \epsilon_2 \epsilon_3 /2}{3-\epsilon_1}.
\end{equation}
There are no approximations in the previous equations.

Next, we focus on the first term of the right hand side of \eqref{masterresta}, i.e $ \overline{ \bra \hat  y_\nk^2 \ket  }$. We define the quantities $Q \equiv \overline{ \bra \hat  y_\nk^2 \ket  } $, $R \equiv \overline{ \bra \hat  p_\nk^2 \ket  } $ and $S \equiv \overline{ \bra \hat  p_\nk \hat  y_\nk  + \hat  y_\nk \hat  p_\nk \ket } $.  The evolution equations of $Q$, $R$, and $S$, can be obtained from the CSL evolution equation \eqref{Htotal}. In fact, for any operator any operator $\hat O$ one has
\begin{equation}\label{vesperadoO}
\frac{d}{d \eta} \overline{\bra \hat O_\nk \ket} = -i  \overline{\bra [ \hat O_\nk, \hat H_\nk    ] \ket }  - \frac{\lambda_k}{2}   \overline{\bra [ \hat y_\nk , [ \hat y_\nk, \hat O_\nk ]] \ket},
\end{equation}
which is the evolution equation of the ensemble average of the expectation value of any operator $\hat O$. Thus, the evolution equations of $Q$, $R$ and $S$ obtained from \eqref{vesperadoO} are:  
\begin{subequations}\label{evolQRS}
	\begin{equation}\label{Qprima}
	Q'= S,
	\end{equation}
	\begin{equation}\label{Rprima}
	R' = -m_1(\eta)S - 2 \overline{m_2(\eta) \bra \hat p_\nk \ket}    + \lambda_k,
	\end{equation}
	\begin{equation}\label{Sprima}
	S' = 2 R - 2 Q m_1(\eta) - 2 \overline{m_2(\eta) \bra \hat y_\nk \ket },
	\end{equation}
\end{subequations}
where
\begin{equation}\label{defm1}
m_1 (\eta) \equiv    k^2 - \frac{a''}{a}  + a^2  V_{, \phi \phi},   
\end{equation}
and
\begin{equation}\label{defm2}
m_2 (\eta) \equiv   - 4 a \phi_0'  \Psi'_\nk + 2 a^3    V_{,  \phi} \Psi_\nk.
\end{equation}

At this point it is important to point out that in our approach the metric perturbation, characterized by $\Psi$, is sourced by $\bra \hat y_\nk \ket$. In particular, that relation is given by our equation \eqref{masterpsi}, which can be rewritten as 
\begin{equation}\label{psiky}
\Psi_\nk = \frac{\phi_0'}{ 2 a M_P^2 \mH} \frac{ \bra \hat y_\nk  \ket}{(1+\epsilon_2)}.
\end{equation}
Therefore, the term $m_2(\eta)$ can be considered as a sort of ``backreaction'' effect of the collapse, since $m_2$ contains explicitly the terms $\Psi'$, $\Psi$. Moreover, by using approximation \eqref{psiprima}, i.e. $\Psi'_{\nk} \simeq \mH \epsilon_2 \Psi_{\nk}$, together with \eqref{psiky}, we reexpress $m_2$ as 
\begin{equation}\label{m2}
m_2 (\eta) \simeq  \bra \hat y_\nk  \ket \mH^2 (-6 \epsilon_1 +2 \epsilon_1^2 +\epsilon_1 \epsilon_2 ).
\end{equation} 
%
From Eq. \eqref{m2}, we see that  $\Psi'$ and $\Psi$ induce terms of order 2 in the HFF.  Using assumptions (i) and (ii) mentioned at the end of Sec. \ref{Sec3} and Eq. \eqref{m2}, we rewrite the evolution equations \eqref{evolQRS} as
\begin{subequations}\label{evolQRS2}
	\begin{equation}\label{Qprima2}
	Q'= S,
	\end{equation}
	\begin{equation}\label{Rprima2}
	R' = -[ k^2 - M(\eta) ]S  + \lambda_k,
	\end{equation}
	\begin{equation}\label{Sprima2}
	S' = 2 R - 2 Q [ k^2 - M(\eta) ],
	\end{equation}
\end{subequations}
where
\begin{equation}\label{defM}
M(\eta) \equiv  \mH^2 \left[ 2- \epsilon_1 + \frac{3}{2} \epsilon_2  + \frac{\epsilon_2^2}{2}  - \frac{7}{2} \epsilon_1 \epsilon_2 + \frac{\epsilon_2 \epsilon_3}{2}    \right].
\end{equation}
The solutions to Eqs. \eqref{evolQRS2}, are
\begin{subequations}\label{solsQRS}
\begin{equation}\label{solQ}
Q = C_1 y_1^2 + C_2 y_2^2 + C_3 y_1 y_2 + Q_p,
\end{equation}
\begin{equation}\label{solR}
R = C_1 y_1'^2 + C_2 y_2'^2 + C_3 y_1' y_2' + R_p,
\end{equation}
\begin{equation}\label{solS}
S = C_1 2y_1 y_1' + C_2 y_2 y_2' + C_3 ( y_1' y_2 +  y_1 y_2' )+ S_p,
\end{equation}
\end{subequations} 
where $y_1$ and $y_2$ are two linearly independent solutions of 
\begin{equation}\label{evoly}
y_{1,2}'' + \left[ k^2 - M (\eta) \right]  y_{1,2} = 0.
\end{equation}
The functions $Q_p$, $R_p$ and $S_p$ are particular solutions of the system \eqref{evolQRS2}. The constants $C_i$, with $i=1,2,3$ are determined by imposing the initial conditions corresponding to the Bunch-Davies vacuum state. The function $Q(\eta)$ is the quantity that we are interested. We proceed  to solve \eqref{evoly}.  

At first order in the HFF, equation \eqref{evoly} is solved exactly in terms of Bessel functions. However, at second order we require new techniques. Here we choose to use the uniform approximation technique \cite{habib2004}. The idea is to rewrite the term  $M(\eta)$ as
\begin{equation}\label{defnu}
M(\eta) = \frac{\nu(\eta)^2 - 1/4   }{\eta^2},
\end{equation}
where the former equation should be understood as the definition of the function $\nu(\eta)$. Then two new functions are introduced:
\begin{equation}\label{defgyf}
g(\eta) \equiv \frac{\nu^2}{\eta^2} - k^2, \qquad f(\eta) \equiv \frac{|\eta-\eta_* |}{\eta-\eta_*} \bigg| \frac{3}{2} \int_{\eta_*}^{\eta} d\tilde \eta \sqrt{g(\tilde \eta)}     \bigg|^{2/3}.
\end{equation}
The time $\eta_*$ is defined by the condition $g(\eta_*)=0$ and is called the turning point, i.e. $\eta_* \equiv -\nu(\eta_*)/k $. According to the uniform approximation, the two linearly independent solutions of \eqref{evoly} are
\begin{equation}\label{key}
y_1 (\eta) = \left(  \frac{f}{g}  \right)^{1/4} \text{Ai} (f), \qquad y_2 (\eta) = \left(  \frac{f}{g}  \right)^{1/4} \text{Bi} (f),
\end{equation}
where Ai and Bi denote the Airy functions of first and second kind respectively. One advantage of the Airy functions is that their asymptotic behavior is quite familiar.

At the onset of inflation, i.e. when $\eta = \tau \to -\infty$, we have
\begin{equation}\label{key}
g^{1/4} = \sqrt{k} e^{i \pi/4}, \qquad f= -\bigg|\frac{3}{2} k \tau \bigg|^{2/3}.
\end{equation} 
In this regime,  the Airy functions oscillate. Specifically, if $x \to +\infty$, then
\begin{equation}\label{airyosc}
\text{Ai}(-x)  \simeq  \frac{ \sin\left( \frac{2}{3} x^{3/2} + \frac{\pi}{4}    \right) }{\sqrt{\pi} x^{1/4} }, \quad \text{Bi}(-x)  \simeq  \frac{ \cos\left( \frac{2}{3} x^{3/2} + \frac{\pi}{4}    \right) }{\sqrt{\pi} x^{1/4} }.
\end{equation}
Thus, at the beginning of inflation we approximate the solutions
\begin{subequations}\label{solsydentro}
	\begin{equation}
y_1(\tau) \simeq \frac{1}{\sqrt{k \pi}} \sin\left( |k \tau | + \frac{\pi}{4}    \right), 	
	\end{equation}
\begin{equation}
 y_2(\tau) \simeq \frac{1}{\sqrt{k \pi}} \cos\left( |k \tau |  + \frac{\pi}{4}    \right).
\end{equation}
\end{subequations}

On the other hand, the Airy functions exhibit exponential behavior for large and positive arguments. That is, for $x \to +\infty$, the Airy functions are approximated by 
\begin{subequations}\label{airyexp}
	\begin{equation}
	\text{Ai}(x)  \simeq \frac{1}{2\sqrt{\pi}}  x^{-1/4} \exp\left( -  \frac{2}{3} x^{3/2} \right),
	\end{equation}
	\begin{equation}
	\text{Bi}(x)  \simeq \frac{1}{\sqrt{\pi}}  x^{-1/4} \exp\left(  \frac{2}{3} x^{3/2}  \right).
	\end{equation}
\end{subequations}

Therefore, in the super-Hubble regime, that is, when $|k \eta| \to 0$, the approximated solutions are
\begin{subequations}\label{solsyfuera}
	\begin{equation}
	y_1 (\eta) \simeq \frac{1}{2\sqrt{\pi}}  g^{-1/4} \exp\left( -  \frac{2}{3} f^{3/2} \right),
	\end{equation}
	\begin{equation}
	y_2 (\eta) \simeq \frac{1}{\sqrt{\pi}}  g^{-1/4} \exp\left(  \frac{2}{3} f^{3/2}  \right).
	\end{equation}
\end{subequations}
Note that in this regime $f \to +\infty$.

Taking into account that the power spectrum is evaluated in the super-Hubble regime, and by considering the exponential solutions \eqref{solsydentro}, together with $g^{1/2} \simeq -\nu/\eta $, we conclude that the term $C_2 y_2^2$ dominates over the rest of the terms in \eqref{solQ}. The quantity of interest is then
\begin{equation}\label{Qfinal}
Q(\eta) \equiv \overline{ \bra \hat  y_\nk^2 \ket  }  \simeq  \frac{C_2}{\pi} \frac{(-\eta)}{\nu} e^{2 F},
\end{equation}
with
\begin{equation}\label{defF}
F \equiv \frac{2}{3} f^{3/2}.
\end{equation}
The constants $C_i$ are found by imposing the initial conditions $Q(\tau) = 1/2k $, $R(\tau) = k/2 $, $S(\tau) = 0 $ and using the approximated solutions \eqref{solsydentro} in the system of equations \eqref{solsQRS}. One also has to take into account the solutions $Q_p$, $R_p$ and $S_p$. In particular, in the sub-Hubble regime, $Q_p \simeq \lambda_k \tau/2k^2$. The constant $C_2$ of \eqref{Qfinal} obtained is
\begin{equation}
C_2 = \frac{\pi}{2} + \frac{\lambda_k \pi  |k \tau|}{2 k^2} + \frac{\lambda_k \pi}{4 k^2} \cos(2 |k \tau|).
\end{equation}
This completes the calculation of $ Q = \overline{\bra \hat y_\nk^2  \ket}$. Now let us focus on the second term on the right hand side of \eqref{masterresta}, i.e. the term $ [4 \text{Re}(A_k)  ]^{-1}$.

 We apply the CSL evolution operator as characterized by Eq. \eqref{Htotal} to the wave function \eqref{psionday}, and regroup terms of order $y^2$, $y^1$ and $y^0$; the evolution equations corresponding to these terms are thus decoupled. Fortunately, the evolution equation corresponding to $y^2$ only contains  $A_k(\eta)$, which is the function we are interested in. The evolution equation is then 
\begin{equation}\label{evolAk}
A_k' = \frac{i}{2} \left[ k^2 - M (\eta) \right]  + \lambda_k - 2iA_k^2, 
\end{equation}
where once again we have assumed that the Newtonian potential is sourced by the expectation value $\bra \hat y_\nk \ket$, and the assumptions (i) and (ii) mentioned at the end of Sec. \ref{Sec3}. By performing the change of variable $A_k \equiv u_k'/(2iu_k)$, the evolution equation of $A_k$ is equivalent to
\begin{equation}\label{evolu}
u_k''  + \left[  q^2 - M(\eta)    \right] u_k = 0,
\end{equation} 
where we have introduced
\begin{equation}\label{defq}
q^2 \equiv k^2 \left( 1 - \frac{2i\lambda_k}{k^2} \right).
\end{equation}

Equation \eqref{evolu} is of the same form as \eqref{evoly}. The general solution is thus
\begin{equation}\label{solu}
u = c_1 \left(  \frac{f}{g}  \right)^{1/4} \text{Ai} (f) + c_2 \left(  \frac{f}{g}  \right)^{1/4} \text{Bi} (f),
\end{equation}
the definitions of $g$ and $f$ given by \eqref{defgyf} hold as before, with the replacement $k^2 \to q^2$ in \eqref{defgyf}. Henceforth,  $g$ and $f$ are complex functions in this case.

The constants $c_{1,2}$ are found by imposing the initial conditions associated to the Bunch-Davies vacuum: $A_k(\tau) = k/2 $. Therefore, by using the asymptotic behavior of the Airy functions when $\eta = \tau \to -\infty$ given by \eqref{airyosc}, we find that
\begin{equation}
c_1 = \sqrt{\frac{\pi}{2}} e^{-i \pi/4}, \qquad c_2 = i c_1. 
\end{equation}

 It is straightforward to check that, 
\begin{equation}\label{ReA}
\text{Re}(A_k)   = \frac{W_k}{|u_k|^2 4i},
\end{equation}
where, $W_k$ is the Wronskian of \eqref{evolu}, i.e. $W_k = u_k' u_k^{*} - u_k^{*'} u_k$. We now proceed to evaluate Re$(A_k)$ in the regime of observational interest, that is, when $-k\eta \to 0$. As before, in this regime, the Airy functions can be approximated by \eqref{airyexp}. Consequently, the solution $u_k$ is 
\begin{equation}\label{soluapp}
u_k(\eta) \simeq \frac{e^{-i \pi 4}}{\sqrt{2} g^{1/4} }  \left[  \frac{ 1 }{2} \exp\left( -  \frac{2}{3} f^{3/2} \right) + i \exp\left(   \frac{2}{3} f^{3/2} \right)   \right].
\end{equation}   
After a long series of calculations using \eqref{ReA}, \eqref{soluapp} and $g^{1/2} \simeq -\nu/\eta$, we find
\begin{equation}\label{inv4reA}
\frac{1}{4 \text{Re}(A_k) } \simeq \frac{-\eta}{2 \nu} \frac{e^{2 \text{Re}(\mathcal{F})}}{\cos [2 \text{Im}(\mathcal{F})]},
\end{equation}
where $\mathcal{F} \equiv \frac{2}{3} f^{3/2}  $. In principle $\mathcal{F} \neq F$, although their definition in terms of $f$ is the same (see \eqref{defF}), the quantity $F$ is real and $\mathcal{F}$ is complex.

Putting together $Q$ given in \eqref{Qfinal}, and  $[4 \text{Re}(A_k) ]^{-1}$ obtained in \eqref{inv4reA}, we can finally obtain $ \overline{ \bra \hat  y_\nk \ket^2  } = Q - [4 \textrm{Re}(A_k)]^{-1}   $, which is 
\begin{eqnarray}\label{masterycuadrado}
 \overline{ \bra \hat  y_\nk \ket^2  } &=& \frac{-\eta}{2 \nu} \bigg[  \left( 1 + \frac{\lambda_k   |k \tau|}{ k^2} + \frac{\lambda_k }{2 k^2} \cos(2 |k \tau|)  \right)  e^{2 F} \nn 
 & -& \frac{e^{2 \text{Re}(\mathcal{F})}}{\cos [2 \text{Im}(\mathcal{F})]}  \bigg]. 
\end{eqnarray}
The last equation is the main result of this Appendix. 

\section{Calculation of the scalar power spectrum at second order}\label{app:calculationPs}

In this Appendix, we proceed to compute the explicit form of the scalar power spectrum at second order in the HFF. 

Using our previous main results, Eqs. \eqref{masterPS}, \eqref{masterpromedios} and \eqref{masterycuadrado}, the full power spectrum is
\begin{eqnarray}\label{PS0}
\mP_{s} &=& \frac{ k^3}{4 \pi^2 M_P^2}  \frac{(1+ \epsilon_1 + \epsilon_2)^2}{(1+\epsilon_2)^2} \frac{|\eta| e^{2 F} }{a^2 \epsilon_1 \nu} \nn 
&\times& \bigg[  \left( 1 + \frac{\lambda_k   |k \tau|}{ k^2} + \frac{\lambda_k }{2 k^2} \cos(2 |k \tau|)  \right) \nn 
& -& \frac{\exp\{2 [\text{Re}(\mathcal{F}) - F]\}  }{\cos [2 \text{Im}(\mathcal{F})]}  \bigg].
\end{eqnarray}

In the former expression, there are functions that depend on $\eta$, these are: $a^2(\eta)$, $\epsilon_{1,2}(\eta)$, $\nu(\eta)$, $F(\eta)$ and $\mathcal{F} (\eta)$. However as we will show in the following, when these functions are expressed explicitly as a function of $\eta$, the $\mP_s$ remains a constant, i.e. independent of $\eta$.  Furthermore, we will express all of these functions at second order in the HFF, and finally exhibit explicitly the $k$ dependence that for now remains implicit in some terms of \eqref{PS0}. This latter step is required to identify the so called spectral index, and running  of the spectral index. In fact, we will make use of some the results obtained in \cite{venninSR} and \cite{ringevalSR}.

We begin by recalling our definition of $\nu$, \eqref{defnu}, which is explicitly given by 
\begin{equation}\label{defnuraiz}
\nu =  \left[  \frac{1}{4} + \eta^2 \mH^2 (2- \epsilon_1 + \frac{3}{2} \epsilon_2  + \frac{\epsilon_2^2}{2}  - \frac{7}{2} \epsilon_1 \epsilon_2 + \frac{\epsilon_2 \epsilon_3}{2}    ) \right]^{1/2}.
\end{equation} 
In that equation, the functions $\mH^2$, $\epsilon_{1,2}$ depend on $\eta$, however the second order terms involving $\epsilon_{1,2,3}$ can be already considered to be constant.

The explicit dependence on the linear terms  $\epsilon_{1,2}$, can be found by expanding around $N_*$. We remind the reader the definition $N \equiv \ln (a/a_{\text{ini}}) $ and that $\eta_*$ represents the turning point i.e. it is the time at which $g(\eta_*) = 0$, see \eqref{defgyf}. Thus, $N_*$ is evaluated at $a(\eta_*)$. The expansion yields
\begin{eqnarray}\label{epsilon1expand}
\epsilon_{1} &=& \epsilon_{1*} +  \frac{d\epsilon_1}{d N} \bigg|_* (N - N_*) + \frac{1}{2} \frac{d^2\epsilon_1}{d {N^2}} \bigg|_* (N - N_*)^2 + \ldots \nn
&=&  \epsilon_{1*} +  \epsilon_{1*}  \epsilon_{2*} \ln \left( \frac{a}{a_*}   \right) \nn 
& +& \frac{1}{2} \left(\epsilon_{1*} \epsilon_{2*}^2 + \epsilon_{1*} \epsilon_{2*} \epsilon_{3*}  \right) \ln^2 \left( \frac{a}{a_*}   \right) + \ldots \nn
&=&  \epsilon_{1*} -  \epsilon_{1*}  \epsilon_{2*} \ln \left( \frac{\eta}{\eta_*}    \right) \nn
& +& \frac{1}{2} \left(\epsilon_{1*} \epsilon_{2*}^2 + \epsilon_{1*} \epsilon_{2*} \epsilon_{3*}  \right) \ln^2 \left( \frac{\eta}{\eta_*}    \right) + \ldots
\end{eqnarray}
where in the second line  the definition of the HFF \eqref{defepsilonn} was used, and in the third line we used that $a(\eta) \propto 1/\eta$. 
As we see from \eqref{epsilon1expand}, the function $\epsilon_1$ now exhibits explicit its $\eta$ dependence.  A similar procedure is used to obtain
\begin{eqnarray}\label{epsilon2expand}
\epsilon_2 &=& \epsilon_{2*} -  \epsilon_{2*}  \epsilon_{3*} \ln \left( \frac{\eta}{\eta_*}    \right) \nn 
&+& \frac{1}{2} \left(\epsilon_{2*} \epsilon_{3*}^2 + \epsilon_{2*} \epsilon_{3*} \epsilon_{4*}  \right) \ln^2 \left(  \frac{\eta}{\eta_*}    \right) + \ldots
\end{eqnarray} 

Using the expansions \eqref{epsilon1expand} and \eqref{epsilon2expand}, one can find the expression for $\mH$ up to second order in HFF, this is \cite{venninSR}:
\begin{equation}\label{Hexpand}
\mH= \frac{-1}{\eta} \left( 1 + \epsilon_{1*} + \epsilon_{1*}^2 + \epsilon_{1*} \epsilon_{2*}       \right) + \epsilon_{1*}  \epsilon_{2*} \frac{1}{\eta} \ln \left( \frac{\eta}{\eta_*}  \right) + \mO(\epsilon^3).
\end{equation}
Moreover, from the last equation one can find an expression for $a$ expanded at second order in HFF \cite{venninSR}
\begin{eqnarray}\label{afinal}
a(\eta) &\simeq& \frac{-1}{H_* \eta}  \bigg[   1 + \epsilon_{1*} + \epsilon_{1*}^2 + \epsilon_{1*} \epsilon_{2*}  \nn 
&-& \left( \epsilon_{1*}  + 2 \epsilon_{1*}^2 + \epsilon_{1*} \epsilon_{2*}  \right)  \ln \left( \frac{\eta}{\eta_*}  \right)   \nn
&+& \frac{1}{2} \left( \epsilon_{1*}^2 + \epsilon_{1*} \epsilon_{2*}   \right)  \ln^2 \left( \frac{\eta}{\eta_*}  \right)  \bigg].
\end{eqnarray}

Therefore, using expansions \eqref{epsilon1expand}, \eqref{epsilon2expand} and \eqref{Hexpand}, the expression corresponding to $\nu$ \eqref{defnuraiz} expanded up to second order in HFF is
\begin{equation}\label{nufinal}
\nu(\eta) = \nu_* - \left(  \epsilon_{1*}  \epsilon_{2*} + \frac{1}{2}  \epsilon_{2*}  \epsilon_{3*} \right)  \ln \left( \frac{\eta}{\eta_*}  \right)  + \mO(\epsilon^3)
\end{equation}
with
\begin{equation}\label{nuasterisco}
\nu_* \equiv \frac{3}{2} + \epsilon_{1*} +  \epsilon_{1*}^2 + \frac{1}{2}  \epsilon_{2*} + \frac{5}{6}  \epsilon_{1*}  \epsilon_{2*} +  \frac{1}{6}  \epsilon_{2*}  \epsilon_{3*}.
\end{equation}

At this point, we have found the explicit dependence in the $\eta$ variable corresponding to the functions: $\epsilon_{1,2} (\eta)$,  $a(\eta)$ and $\nu (\eta)$. But we still require to calculate the functions $\mathcal{F}$ and $F$ to obtain the complete expression for $\mP_s$ \eqref{PS0}. This will be done by solving the corresponding integrals.

 Let us focus on $\mF$. From the definition $\mF\equiv \frac{2}{3} f^{3/2}$ and  $f$,$g$, defined in \eqref{defgyf}, we have
\begin{equation}\label{defintegralmF}
\mF = \int_{\eta_*}^\eta  d\tilde \eta \: \: \sqrt{  \frac{\nu^2 (\tilde \eta )}{\tilde \eta^2 }  - q^2      }.
\end{equation}
Using the definition of $q^2$ \eqref{defq}, we can check that if $\lambda_k =0$, i.e. if there is no collapse of the wave function, then $q^2 = k^2$. Thus, $\mF = F$ when $\lambda_k=0$  (recall $F$ is defined in \eqref{defF}, and that $F$ is real while $\mF$ is complex). Therefore, we can obtain $\mF$ and $F$ from the same integral, i.e. solving integral \eqref{defintegralmF},  automatically yields $\mF$, and by setting $ \lambda_k=0$ in that result, we can obtain also $F$.  

Inserting \eqref{nufinal} into the previous formula and expanding everything to second order, the integrand in \eqref{defintegralmF} reads
\begin{eqnarray}
\sqrt{  \frac{\nu^2 (\tilde \eta )}{\tilde \eta^2 }  - q^2      } &\simeq& \frac{- \nu_*}{\tilde \eta } \left( 1- \frac{q^2 \tilde \eta^2}{\nu_*^2}  \right)^{1/2} \nn 
&+& \frac{3}{2 \nu_*   \tilde \eta } \left(  \epsilon_{1*}  \epsilon_{2*} + \frac{1}{2}  \epsilon_{2*}  \epsilon_{3*}  \right)  \nn
&\times& \left( 1- \frac{q^2 \tilde \eta^2}{\nu_*^2}  \right)^{-1/2}    \ln \left( \frac{\tilde \eta}{\eta_*}  \right).
\end{eqnarray}
Therefore, we have two different integrals to calculate in order to evaluate the term $\mF$. In the following we write,
\begin{equation}
\mF \equiv \mF_1 + \mF_2,
\end{equation}
and calculate each of the $\mF_{1,2}$ separately. These integrals can be solved analytically \cite{venninSR},  for our model, the result is
\begin{equation}\label{mF1}
\lim_{|\eta| \to 0 }  \mF_1 = -\nu_*\left[  1 +   \ln \left| \frac{\eta}{\eta_*}  \right| - \ln 2 + \ln \zeta_k + i \theta_k \right],
\end{equation}
where we define
\begin{equation}\label{defzetakythetak}
\zeta_k \equiv \left( 1 + \frac{4 \lambda_k^2}{k^4}   \right)^{1/4}, \qquad \theta_k \equiv - \frac{1}{2} \arctan \left( \frac{2 \lambda_k}{k^2} \right)
\end{equation}
and
\begin{widetext}
\begin{eqnarray}\label{mF2}
\lim_{|\eta| \to 0 }  \mF_2 &=& \frac{3}{16 \nu_*}  \left(  \epsilon_{1*}  \epsilon_{2*} + \frac{1}{2}  \epsilon_{2*}  \epsilon_{3*}  \right) \left( 4 \ln^2 \left| \frac{\eta}{\eta_*}  \right| - 4 \ln^2 2 + \frac{\pi^2}{3}  \right) + \frac{12}{16 \nu_*} \left(  \epsilon_{1*}  \epsilon_{2*} + \frac{1}{2}  \epsilon_{2*}  \epsilon_{3*}  \right)   \left[ \ln^2 \zeta_k - \theta_k^2  \right]  \nn
&+& \frac{24 i}{16 \nu_*}  \left(  \epsilon_{1*}  \epsilon_{2*} + \frac{1}{2}  \epsilon_{2*}  \epsilon_{3*}  \right) \left( \theta_k \ln \zeta_k  \right) \left[ 1 + \ln \left| \frac{\eta}{\eta_*}  \right| \right].
\end{eqnarray}
\end{widetext}
The explicit dependence on the collapse parameter $\lambda_k$ is now manifested in the previous equations through $\zeta_k$ and $\theta_k$. We notice that $\mF_1$ and $\mF_2$ contain terms that are logarithmically divergent in the limit $|\eta| \to 0$. We will see that this is not a serious problem, the final expression of $\mP_s(k)$ will not have any divergent terms. 

Equations \eqref{mF1} and \eqref{mF2} enable us to calculate $\mF \equiv \mF_1 + \mF_2$ and $F$. The latter, as we have indicated previously, is obtained by setting $\lambda_k =0$, i.e. $\zeta_k =1$ and $\theta_k = 0$, yielding
\begin{eqnarray}\label{Ffinal}
F &=& -\nu_*\left[  1 +   \ln \left| \frac{\eta}{\eta_*}  \right| - \ln 2 \right] + \frac{3}{16 \nu_*}  \left(  \epsilon_{1*}  \epsilon_{2*} + \frac{1}{2}  \epsilon_{2*}  \epsilon_{3*}  \right) \nn
&\times& \left( 4 \ln^2 \left| \frac{\eta}{\eta_*}  \right| - 4 \ln^2 2 + \frac{\pi^2}{3}  \right).
\end{eqnarray}

Additionally, from the resulting expression of $ \mF_1 + \mF_2 = \mF$ we have,
\begin{eqnarray}\label{ImF}
2 \text{Im}(\mathcal{F}) &=& - 2 \nu_* \theta_k + \frac{3}{\nu_*}  \left(  \epsilon_{1*}  \epsilon_{2*} + \frac{1}{2}  \epsilon_{2*}  \epsilon_{3*}  \right) \left( \theta_k \ln \zeta_k  \right) \nn
 &\times& \left[ 1 + \ln \left| \frac{\eta}{\eta_*}  \right| \right],
\end{eqnarray}

\begin{widetext}
and
\begin{equation}\label{ReF}
2 [\text{Re}(\mathcal{F}) - F] = -2 \nu_* \ln \zeta_k + \frac{3}{2 \nu_*} \left(  \epsilon_{1*}  \epsilon_{2*} + \frac{1}{2}  \epsilon_{2*}  \epsilon_{3*}  \right)  \left[ \ln^2 \zeta_k - \theta_k^2  \right].
\end{equation}

We have all the expressions needed to give an expression of the $\mP_s$ in terms of: the collapse parameter and the second order HFF. Therefore, collecting Eqs. \eqref{Ffinal},  \eqref{ImF}, \eqref{ReF}, as well as the corresponding ones to $\epsilon_{1,2} (\eta)$, $a(\eta)$ and $\nu(\eta)$ [this is Eqs. \eqref{epsilon1expand}, \eqref{epsilon2expand}, \eqref{afinal} and \eqref{nufinal}], it is straightforward, although lengthy, to obtain the power spectrum from \eqref{PS0}. The final expression is
 \begin{eqnarray}\label{PS1}
\mP_s &\simeq&  \frac{ 18 e^{-3} H_*^2}{ \pi^2 M_P^2 \epsilon_{1*}} \bigg\{ 1 + \epsilon_{1*} \left( - \frac{2}{3 } + 2 \ln 2 \right) + \epsilon_{2*} \left( - \frac{1}{3} + \ln 2 \right)  \nn
 &+&  \epsilon_{1*}^2 \left( -\frac{26}{9} + \frac{2}{3} \ln 2 + 2 \ln^2 2 \right) + \epsilon_{2*}^2 \left( -\frac{1}{18} - \frac{1}{3} \ln 2 + \frac{1}{2} \ln^2 2 \right)  \nn
 &+& \epsilon_{1*} \epsilon_{2*}  \left( -\frac{43}{9}  + \frac{\pi^2}{12 } +  \frac{1}{3} \ln 2 +  \ln^2 2 \right)  \nn
 &+& \epsilon_{2*} \epsilon_{3*}  \left( -\frac{1}{9}  + \frac{\pi^2}{24} +  \frac{1}{3} \ln 2 -\frac{1}{2}  \ln^2 2 \right) \bigg\} C(k),
 \end{eqnarray}
where we have defined
\begin{eqnarray}\label{defCK}
& & C(k) \equiv   \left( 1 + \frac{\lambda_k   |k \tau|}{ k^2} + \frac{\lambda_k }{2 k^2} \cos(2 |k \tau|)  \right)  - \frac{\exp\{2 [\text{Re}(\mathcal{F}) - F]\}  }{\cos [2 \text{Im}(\mathcal{F})]},
\end{eqnarray}
with
\begin{equation}
\exp \{ 2 [\text{Re}(\mathcal{F}) - F] \} =  \zeta_k^{-3}  \exp \left[ -2   \left(   \epsilon_{1*} +  \epsilon_{1*}^2 + \frac{1}{2}  \epsilon_{2*} + \frac{5}{6}  \epsilon_{1*}  \epsilon_{2*} +  \frac{1}{6}  \epsilon_{2*} \epsilon_{3*}  \right) \ln \zeta_k +  \left(  \epsilon_{1*}  \epsilon_{2*} + \frac{1}{2}  \epsilon_{2*}  \epsilon_{3*}  \right)  \left( \ln^2 \zeta_k - \theta_k^2  \right) \right] 
\end{equation}
and
\begin{eqnarray}
\cos [2 \text{Im}(\mathcal{F})] &=& \cos \bigg\{   \left( 3 + 2 \epsilon_{1*} +  2 \epsilon_{1*}^2 +   \epsilon_{2*} + \frac{5}{3}  \epsilon_{1*}  \epsilon_{2*} +  \frac{1}{3}  \epsilon_{2*}  \epsilon_{3*}  \right)  \theta_k  \nn 
&-& 2  \left(  \epsilon_{1*}  \epsilon_{2*} + \frac{1}{2}  \epsilon_{2*}  \epsilon_{3*}  \right) \left( \theta_k \ln \zeta_k  \right) \left[ 1 + \ln \left| \frac{\eta}{\eta_*}  \right| \right]  \bigg\}. 
\end{eqnarray}
\end{widetext}

At this point a few comments are in order. First, as discussed in Refs. \cite{jmartinWKB,ringevalSR,venninSR}, the presence of the factor $18e^{-3} \simeq 0.896$ is typical for the uniform approximation, and from now on, we will simply set this factor equal to one. Second, the divergent logarithmic term appears only in  $C(k)$ but as an argument of a cosine function, which in turn appears in the denominator in the definition of $C(k)$; thus it represents no problem at all. In fact, we can set $\ln \left| \frac{\eta}{\eta_*}  \right| \simeq \ln \left| \frac{a_*}{a}  \right| = \Delta N_*$, i.e. is the number of e-folds from $\eta_*$ to the end of inflation.  

Our expression of $\mP(k)$ depends on $\eta_*$, and the  HFF, as well as $H_*$ all evaluated at $\eta_*$, which is the turning point of $g$, this means $\eta_* \equiv -\nu(\eta_*)/k $. Thus, there is a $k$ dependence that remains hidden in those quantities. In order to uncover the $k$ dependence, we define a pivot wave number $k_\di$ and expand all those terms around an unique conformal time $\eta_\di$. It is customary to set this $\eta_\di$ as the time of ``horizon crossing,"  which is defined as
\begin{equation}
-k_\di \eta_\di = 1.
\end{equation} 
Technically, this means that, for instance, the Hubble parameter $H_*$ must be rewritten as an expansion around $\eta_\di$,
\begin{equation}\label{Hx}
H_* = H_\di \left[   1 + \left( \epsilon_{1 \di}^2 + \epsilon_{1 \di}  \right) \ln \frac{\eta_*}{\eta_\di}    + \frac{1}{2} \left(\epsilon_{1 \di}^2 + \epsilon_{1 \di} \epsilon_{2 \di} \right) \ln^2 \frac{\eta_*}{\eta_\di}     \right]
\end{equation}
and the $k$ dependence is thus uncovered by the relation
\begin{equation}
\frac{\eta_*}{\eta_\di} = \frac{k_\di}{k} \nu_*.   
\end{equation}
Expanding the previous equation at second order in HFF, we obtain 
\begin{eqnarray}\label{logetaxetadi}
\ln \left( \frac{\eta_*}{\eta_\diamond} \right) &=& \left( \ln \frac{3}{2} + \ln \frac{k_\di}{k} \right) \left( 1 - \frac{2}{3} \epsilon_{1 \di}  \epsilon_{2 \di} - \frac{1}{3} \epsilon_{2 \di} \epsilon_{3 \di}   \right) \nn
&+& \frac{2}{3} \epsilon_{1 \di} + \frac{1}{3} \epsilon_{2 \di} + \frac{4}{9} \epsilon_{1 \di}^2  - \frac{1}{18} \epsilon_{2 \di}^2 \nn
 &+& \frac{1}{9} \epsilon_{2 \di} \epsilon_{3 \di} + \frac{1}{3} \epsilon_{1 \di} \epsilon_{2 \di}.   
\end{eqnarray}
Hence, substituting \eqref{logetaxetadi} into \eqref{Hx}, will exhibit explicitly the $k$ dependence in the Hubble factor $H_*$.

Applying this same technique to the HFF $\epsilon_{i *}$ and $\nu_*$ lead us to our main expression. This is, the scalar power spectrum at second order in the HFF given by the CSL model is 

\begin{widetext} 

\begin{eqnarray}\label{PSdiamond}
\mP_s &=& \frac{  H_\di^2}{ \pi^2 M_P^2 \epsilon_{1\di}}  \bigg\{ 1 - D \epsilon_{1\di} - D \epsilon_{2\di} + \left( -\frac{10}{9} - 2D + 2D^2 \right) \epsilon_{1\di}^2  + \left(  \frac{2}{9}  + \frac{D^2}{2}   \right)\epsilon_{2\di}^2                 \nn 
&+& \left(  \frac{-29}{9} - D +D^2 + \frac{\pi^2}{12}          \right) \epsilon_{1\di}\epsilon_{2\di} + \left( \frac{\pi^2}{24} - \frac{1}{18} - \frac{D^2}{2 }\right)\epsilon_{2\di}\epsilon_{3\di}  \nn
&+& \left[    -2 \epsilon_{1\di} - \epsilon_{2\di}+ 2 (-1 + 2D) \epsilon_{1\di}^2 -(1-2D) \epsilon_{1\di} \epsilon_{2\di} + D \epsilon_{2\di}^2 - D \epsilon_{2\di} \epsilon_{3\di}                \right] \ln \left(  \frac{k}{k_\di}  \right) \nn
&+& \left[  2 \epsilon_{1\di}^2 + \epsilon_{1\di} \epsilon_{2\di}  + \frac{1}{2} \epsilon_{2\di}^2  - \frac{1}{2} \epsilon_{2\di} \epsilon_{3\di}    \right] \ln^2 \left(  \frac{k}{k_\di}  \right) \bigg\} C(k),
\end{eqnarray}
where $D \equiv 1/3 - \ln 3 $ and  in $C(k)$ (defined in \eqref{defCK}) we have the following expressions for the $\exp\{2 [\text{Re}(\mathcal{F}) - F]\}/\cos [2 \text{Im}(\mathcal{F})]$ term:
\begin{eqnarray}
\exp \{ 2 [\text{Re}(\mathcal{F}) - F] \} &=&  \zeta_k^{-3}  \exp \bigg\{   \bigg[ -2 \epsilon_{1\di} - \epsilon_{2\di} - 2 \epsilon_{1\di}^2 + \left(  - \frac{5}{3} + 2 \ln \frac{3}{2} - 2 \ln \frac{k}{k_\di} \right) \epsilon_{1\di}  \epsilon_{2\di}  \nn 
&+& \left(  - \frac{1}{3} + \ln \frac{3}{2} -  \ln \frac{k}{k_\di} \right) \epsilon_{2\di}  \epsilon_{3\di}     \bigg]  \ln \zeta_k +   \left(  \epsilon_{1\di}  \epsilon_{2\di} + \frac{1}{2}  \epsilon_{2\di}  \epsilon_{3\di}  \right)  \left( \ln^2 \zeta_k - \theta_k^2  \right)   \bigg\}
\end{eqnarray}
and
\begin{eqnarray}
\cos [2 \text{Im}(\mathcal{F})] &=& \cos \bigg\{   \bigg( -3  -2 \epsilon_{1\di} - \epsilon_{2\di} - 2 \epsilon_{1\di}^2 + \left(  - \frac{5}{3} + 2 \ln \frac{3}{2} - 2 \ln \frac{k}{k_\di} \right) \epsilon_{1\di}  \epsilon_{2\di}  \nn 
&+& \left(  - \frac{1}{3} + \ln \frac{3}{2} -  \ln \frac{k}{k_\di} \right) \epsilon_{2\di}  \epsilon_{3\di}       \bigg)  \theta_k   \nn
&+& 2  \left(  \epsilon_{1\di}  \epsilon_{2\di} + \frac{1}{2}  \epsilon_{2\di}  \epsilon_{3\di}  \right) \left( \theta_k \ln \zeta_k  \right) \left[ 1  -\Delta N_\di - \ln \frac{3}{2} + \ln \frac{k}{k_\di} \right]  \bigg\}.
\end{eqnarray}
Notice that the former divergent logarithmic term, has now transformed into $\Delta N_\di$ which is the number of e-folds from the horizon crossing of the pivot scale $k_\di$ to the end of inflation. Typically  $\Delta N_\di \sim 60$. Equation \eqref{PSdiamond}, is our final expression for the PS, within the CSLIM, written in terms of the HFF and the collapse parameter $\lambda_k$.

In the standard approach for the predicted power spectrum (PS), the $k$  dependence can be parameterized by the so called scalar spectral index $n_s$ and the running of the spectral index $\alpha_s$.  The parameters $n_s$ and $\alpha_s$ are of interest since they are used to constrain the shape of the PS consistent with the observational data.  On the other hand, in our main result \eqref{PSdiamond}, we can see that the CSL model induces an extra $k$ dependence on the PS through $C(k)$ as expected. Consequently, it would be helpful to identify the parameters $n_s$ and $\alpha_s$, and then including them, if necessary, in the function $C(k)$. This will allow us to compare directly the observational consequences between our approach and the standard inflationary model. That preliminary analysis is done in Sec. 4.

Thus, in order to deduce an expression for $n_s$ and $\alpha_s$ in terms of the HFF, we set $C(k)=1$ and follow the method in \cite{ringevalSR}. Additionally, since the amplitude of the PS was computed up to second order in HFF, the expression for $n_s$ is valid up to third order and $\alpha_s$ up to fourth order. Hence, the scalar spectral index is given by
\begin{eqnarray}\label{nscsl}
n_s - 1 &=& - 2 \epsilon_{1\di} - \epsilon_{2\di} - 2 \epsilon_{1\di}^2  - (1+D) \epsilon_{1\di} \epsilon_{2\di} - D \epsilon_{2\di} \epsilon_{3\di} - 2 \epsilon_{1\di}^3 + \left( - \frac{47}{9} -5 D + 3 D^2   \right) \epsilon_{1\di}^2 \epsilon_{2\di}  \nn
&+& \frac{4}{9} \epsilon_{2\di}^2 \epsilon_{3\di} + \left( - \frac{1}{18} - \frac{D^2}{2} + \dfrac{\pi^2}{24}  \right) \epsilon_{2\di} \epsilon_{3\di}^2 + \left( - \frac{1}{18} - \frac{D^2}{2} + \dfrac{\pi^2}{24}   \right) \epsilon_{2\di} \epsilon_{3\di} \epsilon_{4\di} \nn
&+& \left( - \frac{29}{9} - 2D + \frac{\pi^2}{12}\right) \epsilon_{1\di} \epsilon_{2\di} \epsilon_{3\di} + \left( - \frac{38}{9} - D + \frac{\pi^2}{12} \right) \epsilon_{1\di} \epsilon_{2\di}^2
\end{eqnarray}
and the running of the spectral index yields
\begin{eqnarray}\label{alphacsl}
\alpha_s &=& -2 \epsilon_{1\di} \epsilon_{2\di} -\epsilon_{2\di} \epsilon_{3\di} -6 \epsilon_{2\di} \epsilon_{1\di}^2-12 \epsilon_{2\di} \epsilon_{1\di}^3-D \epsilon_{2\di} \epsilon_{3\di}^2-D \epsilon_{2\di} \epsilon_{3\di} \epsilon_{4\di}+\frac{8}{9} \epsilon_{2\di}^2 \epsilon_{3\di}^2+\frac{4}{9} \epsilon_{2\di}^2 \epsilon_{3\di} \epsilon_{4\di}   \nn
&+&  \left(6 D^2-11 D-\frac{121}{9}\right) \epsilon_{2\di}^2 \epsilon_{1\di}^2+\left(3 D^2-6 D-\frac{65}{9}\right) \epsilon_{2\di} \epsilon_{3\di} \epsilon_{1\di}^2+\left(-\frac{D^2}{2}+\frac{\pi ^2}{24}-\frac{1}{18}\right) \epsilon_{2\di} \epsilon_{3\di}^3 \nn
&+& \left(-\frac{D^2}{2}+\frac{\pi ^2}{24}-\frac{1}{18}\right) \epsilon_{2\di} \epsilon_{3\di} \epsilon_{4\di}^2+\left(-\frac{3 D^2}{2}+\frac{\pi ^2}{8}-\frac{1}{6}\right) \epsilon_{2\di} \epsilon_{3\di}^2 \epsilon_{4\di}+\left(-\frac{D^2}{2}+\frac{\pi ^2}{24}-\frac{1}{18}\right) \epsilon_{2\di} \epsilon_{3\di} \epsilon_{4\di} \epsilon_{5\di} \nn 
&+& \left(-D+\frac{\pi ^2}{12}-\frac{38}{9}\right) \epsilon_{2\di}^3 \epsilon_{1\di}+(-D-1) \epsilon_{2\di}^2 \epsilon_{1\di}+\left(-3 D+\frac{\pi ^2}{12}-\frac{29}{9}\right) \epsilon_{2\di} \epsilon_{3\di}^2 \epsilon_{1\di}+\left(-4 D+\frac{\pi ^2}{4}-\frac{38}{3}\right) \epsilon_{2\di}^2 \epsilon_{3\di} \epsilon_{1\di} \nn 
&-& (D+2) \epsilon_{2\di} \epsilon_{3\di} \epsilon_{1\di}+\left(-3 D+\frac{\pi ^2}{12}-\frac{29}{9}\right) \epsilon_{2\di} \epsilon_{3\di} \epsilon _4 \epsilon_{1\di}.
\end{eqnarray}
We note that at the lowest order in the HFF, $n_s$ and $\alpha_s$ coincides with the standard expressions.

\end{widetext}

\bibliography{bibliography}
\bibliographystyle{apsrev}

\end{document}